\documentclass[%
 reprint,
 superscriptaddress,
 amsmath,amssymb,
 aps]{revtex4-2}

\usepackage{graphicx}
\usepackage{dcolumn}
\usepackage{xcolor}
\usepackage[caption=false]{subfig}
\usepackage{bm}
\usepackage{hyperref}
\usepackage[mathlines]{lineno}
\usepackage{float}

\begin{document}

\preprint{APS/123-QED}

\title{Ordering dynamics and aging in the Symmetrical Threshold model}

\author{David Abella}\email{david@ifisc.uib-csic.es}%
\affiliation{%
 Instituto de F\'{\i}sica Interdisciplinar y Sistemas Complejos IFISC (CSIC - UIB), Campus UIB, 07122 Palma de Mallorca, Spain
}%

\author{Juan Carlos González-Avella}%
\affiliation{%
 Instituto de F\'{\i}sica Interdisciplinar y Sistemas Complejos IFISC (CSIC - UIB), Campus UIB, 07122 Palma de Mallorca, Spain
}%
\affiliation{Advanced Programming Solutions SL, 07121 Palma de Mallorca, Spain}

\author{Maxi San Miguel}%
\affiliation{%
 Instituto de F\'{\i}sica Interdisciplinar y Sistemas Complejos IFISC (CSIC - UIB), Campus UIB, 07122 Palma de Mallorca, Spain
}%

\author{Jos\'e J. Ramasco}
\affiliation{%
 Instituto de F\'{\i}sica Interdisciplinar y Sistemas Complejos IFISC (CSIC - UIB), Campus UIB, 07122 Palma de Mallorca, Spain
}%

\date{\today}%

\begin{abstract}
The so-called Granovetter-Watts model was introduced to capture a situation in which the adoption of new ideas or technologies requires a certain redundancy in the social environment of each agent to take effect. This model has become a paradigm for complex contagion. Here we investigate a symmetric version of the model: agents may be in two states that can spread equally through the system via complex contagion. We find three possible phases: a mixed one (dynamically active disordered state), an ordered one, and a heterogeneous frozen phase. These phases exist for several configurations of the contact network. Then we consider the effect of introducing aging as a non-Markovian mechanism in the model, where agents become increasingly resistant to change their state the longer they remain in it.  We show that when aging is present, the mixed phase is replaced, for sparse networks, by a new phase with different dynamical properties. This new phase is characterized by an initial disordering stage followed by a slow ordering process towards a fully ordered absorbing state. In the ordered phase, aging modifies the dynamical properties. For random contact networks, we develop a theoretical description based on an Approximate Master Equation that describes with good accuracy the results of numerical simulations for the model with and without aging.
\end{abstract}

\maketitle


\section{\label{sec:Introduction} Introduction}

A variety of collective phenomena can be well understood through stochastic binary-state models for interacting agents. In these models, each agent is assumed to be in one of two possible states, such as susceptible/infected, adopters/non-adopters, etc., depending on the context of the model. The interaction among agents is determined by the underlying contact network and the dynamical rules of the model. There are various examples of binary-state models, including processes of opinion formation \cite{Voter-original, sood-2005, Suchecki-2005, fernandez-gracia-2014, redner-2019} and disease or social contagion \cite{granovetter-1978, pastor-satorras-2015}, among others. The consensus problem consists in determining under which circumstances the agents end up sharing the same state or when the coexistence of both states prevails. This is characterized by a phase diagram that provides the boundaries separating domains of different behaviors in the control parameter space. Macroscopic descriptions of these models in terms of mean-field, pair, and higher-order approximations are  well established \cite{gleeson-2011}. 

An important category of binary-state models are threshold models \cite{watts-2002}, which were originally introduced by M. Granovetter \cite{granovetter-1978} to address problems of social contagion such as rumor propagation, innovation adoption, riot participation, etc. Multiple exposures, or group interaction, are necessary in these models to update the current state, a characteristic of complex contagion models \cite{centola-2007,unknown-author-2018}. The threshold model presents a discontinuous phase transition from a ``global cascade'' phase to a ``no cascade'' phase, which was analyzed in detail in Ref. \cite{watts-2002}. This model has been extensively studied on various network topologies, such as regular lattices, small-world \cite{centola-2007}, random \cite{gleeson-2007}, clustered \cite{hackett-2011,hackett-2013}, modular \cite{gleeson-2008}, hypergraphs \cite{de-arruda-2020}, homophilic \cite{diaz-diaz-2022} and coevolving \cite{min2023threshold} networks. 

A main difference between the threshold model and other binary-state models, such as the Voter \cite{Voter-original}, majority vote (MV) \cite{de1992isotropic,pereira2005majority,campos2003small}, and nonlinear Voter model \cite{castellano-2009,mobilia2015nonlinear,mellor2016characterization,Min-2017,jewski-2017,peralta-2018}, is the lack of symmetry between the two states. In the threshold model, changing state is only possible in one direction, representing the adoption forever of a new state that initially starts in a small minority of agents. A symmetric version of the threshold model, with possible changes of states in both directions, was introduced in Refs. \cite{nowak2019homogeneous,nowak2020symmetrical} to investigate the impact of noise and anticonformity. However, a complete characterization of the Symmetrical Threshold model and its ordering dynamics have not been addressed so far.

Aging is an important non-Markovian effect in binary-state models that has significant implications. It describes how the persistence time of an agent in a particular state influences the transition rate to a different state \cite{stark-2008, fernandez-gracia-2011, perez-2016, boguna-2014, chen-2020}. As such, the longer an agent remains in the current state, the smaller the probability of changing. Aging has been shown to cause coarsening dynamics towards a consensus state in the Voter model \cite{fernandez-gracia-2011,peralta-2020C}, to induce bona fide continuous phase transitions in the noisy Voter model \cite{artime-2018,peralta-2020A}, modify the phase diagram and non-equilibrium dynamics of the Schelling segregation model \cite{Abella-2022}, and to modify non-trivially the cascade dynamics of the threshold model \cite{Abella-2022-AME}. The introduction of aging is motivated by strong empirical evidence that human interactions do not occur at a constant rate and cannot be described using a Markovian assumption. Empirical studies have reported heavy-tail inter-event time distributions that reflect heterogeneous temporal activity patterns in social interactions \cite{iribarren-2009, karsai-2011, rybski-2012, zignani-2016, artime-2017, kumar-2020}.

In this work, we present a comprehensive analysis of the Symmetrical Threshold model, including its full phase diagram, and we investigate the effects of aging in the model. The model is examined in various network topologies, such as the complete graph, Erd\H{o}s-Rényi (ER)  \cite{erdos1960evolution}, random regular (RR) \cite{wormald1999models}, and a two-dimensional Moore lattice. The possible phases of the system are defined by the final stationary state as well as by the ordering/disordering dynamics characterized by the time-dependent magnetization and interface density, the persistence, and the mean internal time. For both the original model and the aging variant, the results of Monte Carlo numerical simulations are compared with results from the theoretical framework provided by an Approximate Master Equation (AME)\cite{gleeson-2013,Abella-2022-AME}, which is general for any random network. We also derive a mean-field analysis to describe the outcomes in a complete graph.

The article is organized as follows: In Section \ref{Symmetrical Threshold model}, we describe the Symmetrical Threshold model and provide the numerical and theoretical analysis of the phase diagram. Each subsection reports the results for the different networks chosen. Section \ref{Symmetrical Threshold model with aging} presents the Symmetrical Threshold model with aging, the corresponding numerical and theoretical analysis, and the comparison with the model without aging. The results for the Moore lattice are shown in Section \ref{sec: Symmetrical Threshold model in a Moore_Lattice}. Finally, we conclude with a summary and conclusions in Section \ref{Summary and conclusions}.

\section{\label{Symmetrical Threshold model} Symmetrical Threshold model}

\begin{figure}
    \includegraphics[width=\columnwidth]{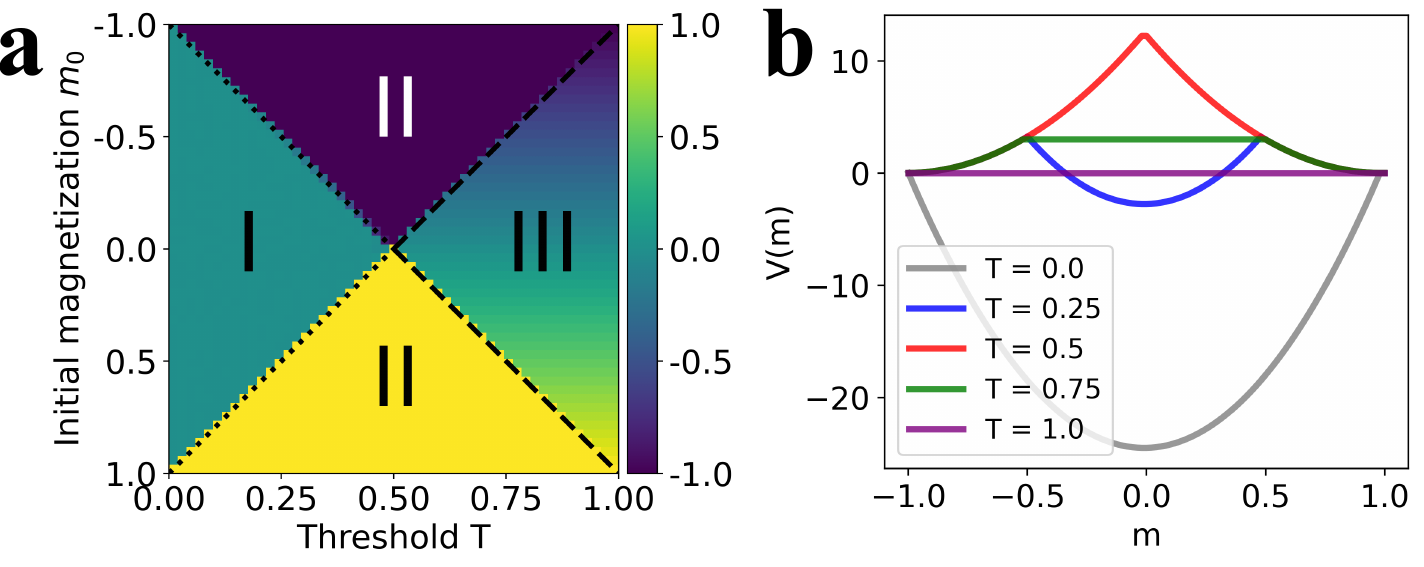}
    \caption{\label{COM_LAT_PD} (a) Phase diagram of the Symmetrical Threshold model in a Complete graph of $N = 2500$ nodes. Dotted and dashed lines correspond to $T = (1-|m_0|)/2$ and $T = (1+|m_0|)/2$, respectively. Average performed over 5000 realizations. (b) Potential representation from Eq. \eqref{eq:potential} for a set of values of the threshold $T$, shown in different colors.}
\end{figure}

The system consists of a set of $N$ agents located at the nodes of a network. The variable describing the state of each agent $i$ takes one of the two possible values: $s_i = \pm 1$. Every agent has assigned a fixed threshold $0 \leq T \leq 1$, which determines the fraction of different neighbors required to change state. Even though this value might be agent dependent, we will consider here only the case with a homogeneous $T$ value for all the agents of the system. In each update attempt, an agent $i$ (called active agent) is randomly selected, and if the fraction of neighbors with a different state is larger than the threshold $T$, the active agent changes state $s_i \to -s_i$. Simulation time is measured in Monte Carlo (MC) steps, i.e., $N$ update attempts. Numerical simulations run until the system reaches a frozen configuration (absorbing state) or until the average magnetization, $m = (1/N) \sum_i s_i$, fluctuates around a constant value.

\subsection{Mean-field}

We first consider the mean-field case of the complete graph (all-to-all connections). We take an initial random configuration with magnetization $m_0$ and perform numerical simulations for various values of $T$ to construct the phase diagram (shown in Fig. \ref{COM_LAT_PD}a). We find three different phases based on the final state:
\begin{itemize}
    \item \textbf{Phase I or Mixed}: The system reaches an active disordered state (final magnetization $m_f = 0$) where the agents change their state continuously;
    \item \textbf{Phase II or Ordered}: The system reaches the ordered absorbing states ($m_f = \pm 1$) according to the initial magnetization $m_0$;
    \item \textbf{Phase III or Frozen}: The system freezes at the initial random state $m_f = m_0$.
\end{itemize}

\begin{figure*}
    \includegraphics[width=\linewidth]{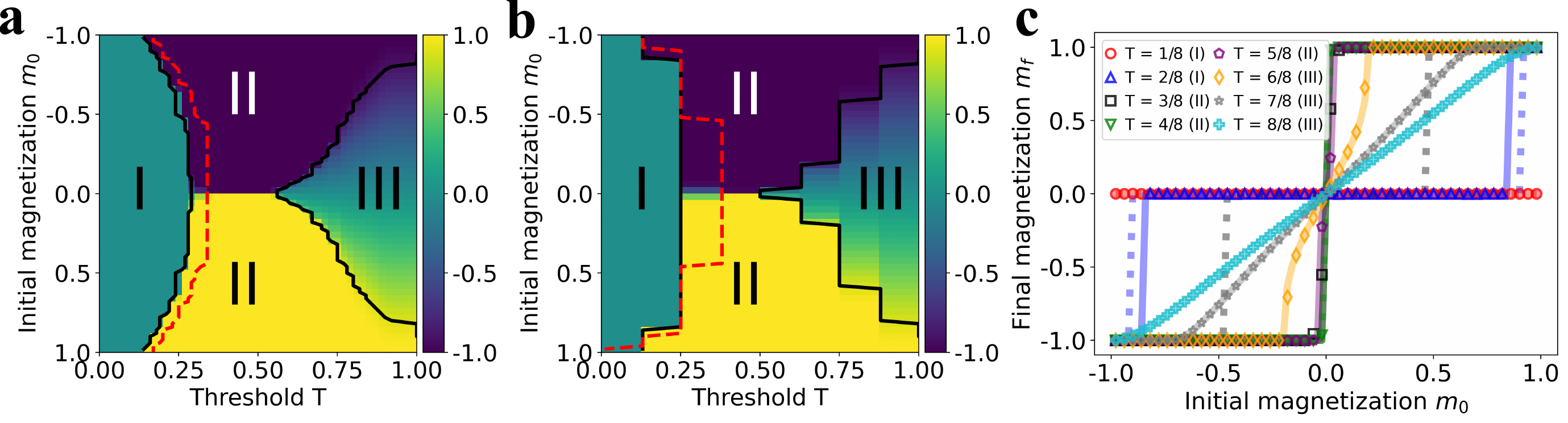}
    \caption{\label{ER_REG_PD} Phase diagram of the Symmetrical Threshold model in an ER (a) and a RR (b) graph, both of $N=4\cdot10^4$ nodes and mean degree $\langle k \rangle=8$. The color map indicates the value of the average final magnetization $m_f$. The red dashed line is the HMF prediction of the mixed-ordered critical line. The black solid lines correspond to the AME prediction of the borders of Phase II. (c) Average final magnetization $m_f$ as a function of the initial magnetization $m_0$ for different $T$ values (indicated with different colors and markers) in the RR graph. The average is performed over 5000 realizations. The dotted and solid lines are the HMF (for $T=1/8 - 4/8$) and AME predictions (for all $T$), respectively.}
\end{figure*}

For a given initial magnetization $m_0 \neq 0$ and increasing $T$, the system undergoes a mixed-ordered transition at a critical threshold $T_{c} = (1-|m_0|)/2$, and an ordered-frozen transition at a critical threshold $T_{c}^{*} = (1 + |m_0|)/2 > T_{c}$ (indicated by dotted and dashed black lines in Fig. \ref{COM_LAT_PD}a). In this mean-field scheme, if the fraction of nodes in state $+1$ is denoted by $x$, the condition for a node in state $-1$ to change its state is given by $\theta(x - T)$, where  $\theta$ is the Heaviside step function. Thus, in the thermodynamic limit ($N\to \infty$), the variable $x$ evolves over time according to the following mean-field equation:
\begin{equation}
    \frac{dx}{dt} = (1 - x) \; \theta(x - T) - x \; \theta(1 - x - T) = - \frac{\partial V(x)}{\partial x}.
\end{equation}
Here, $V(x)$ is the potential function. The stationary value of $x$, $x_{\rm st}$, is the solution of the implicit equation resulting from setting the time derivative equal to 0. No analytical expression of $x_{\rm st}$ has been found in terms of $T$, but the solutions can be understood in terms of the potential $V(x)$:
\begin{align}
    \label{eq:potential}
    V(x) = & - \int (1 - x) \; \theta(x - T) - x \; \theta(1 - x - T) \; dx \nonumber\\
    = & \frac{x^2}{2} + \frac{1}{2} \left( T^2 - 2T - x^2 + 1\right) \; \theta(T+x-1)\nonumber\\
    & - \frac{1}{2} \left( T^2 - 2T - x(x-2)\right) \; \theta(x - T) .
\end{align}
The minimum and maximum values of $V(x)$ correspond to stable and unstable solutions, respectively. Figure \ref{COM_LAT_PD}b shows the potential's dependence on the magnetization, obtained after a variable change $m = 2x-1$ in Eq. \eqref{eq:potential}. For $T < 0.5$, $m = 0$ is a stable solution, but increasing the threshold reduces the range of values of the initial magnetization from which this solution is reached, enclosing Phase I between the unstable solutions $m = 1-2\, T$ and $2\, T-1$. In fact, if $m_0 > 1-2\, T$, the system reaches the absorbing solution $m=+1$, while if $m_0 < -1+2\, T$, it reaches $m=-1$ (Phase II). For $T = 0.5$, there is just one unstable solution at $m=0$, and all the initial magnetization values reach the absorbing states $m=\pm 1$. For $T > 0.5$, the potential is equal to a constant value for a range of $m_0$, which means that an initial condition will remain in this state forever (Phase III). The range of values of the initial condition from which this phase is reached grows linearly with $T$ until $T=1$, where all initial conditions fulfill $\frac{dm}{dt}=0$.

Note that the mean-field Symmetrical Threshold model for $T=1$ shows the same potential profile as the mean-field Voter model \cite{Suchecki-2005, Voter-original, Voter}. The important difference is that for the Voter model, any initial magnetization is marginally stable, while in our model any initial magnetization is an absorbing state in Phase III. In the Voter model finite size fluctuations will take the system to the absorbing states $m=\pm 1$. 


\subsection{Random networks}

We analyze the phase diagram of the Symmetrical Threshold model in two random networks: Erd\H{o}s-Rényi (ER) \cite{erdos1960evolution} and random regular (RR) \cite{wormald1999models} graphs with mean degree $\langle k \rangle = 8$. Figures \ref{ER_REG_PD}a and \ref{ER_REG_PD}b show the phase diagram for both networks, where it is shown that the existence of the three phases previously described is robust to changes in network structure. The main difference from the all-to-all scenario is that Phase III does not freeze exactly at the same initial magnetization. Instead, the system reaches an absorbing state with a higher magnetization $m_f > m_0$. In this phase, the value of $m_f$ depends on the threshold such that increasing $T$, increases the disorder in the system, until $T = 1$, where $m_f = m_0$ (see Fig. \ref{ER_REG_PD}c). On the other hand, phases I and II reach the same stationary state as in the mean-field case. Furthermore, the critical thresholds $T_{c}$ and $T_{c}^{*}$ show a different dependence on $m_0$ depending on the network structure. 

To explain the transitions exhibited by the model, we use a theoretical framework for binary-state dynamics in complex networks \cite{gleeson-2013}: the Approximate Master Equation (AME), which considers agents in both states $\pm 1$ with degree $k$, $m$ neighbors in state $-1$ that have been $j$ time steps in the current state (called ``internal time'' or ``age'') as different sets in a compartmental model (see details of the AME derivation in our previous work \cite{Abella-2022-AME}). In general, the AME is:
\begin{align}
\label{eq:AME_age}
    \frac{d}{d t} x^{\pm}_{k, m, 0}=&- x^{\pm}_{k, m, 0} + \sum_l T^{\mp}_{k, m,l} \, x^{\mp}_{k, m, l} \nonumber\\
    &- \beta^{\pm} \, (k-m) \, x^{\pm}_{k, m, 0} -\gamma^{\pm} \, m \, x^{\mp}_{k, m, 0}, \nonumber\\
    \frac{d}{d t} x^{\pm}_{k, m, j}=&- x^{\pm}_{k, m, j}+ A^{\pm}_{k, m,j} \, x^{\pm}_{k, m, j-1} - \beta^{\pm} \, (k-m) \, x^{\pm}_{k, m, j}\nonumber\\
    &+ \beta^{\pm} \, (k-m+1) \, x^{\pm}_{k, m-1, j-1}\\
    & +\gamma^{\pm} \, (m+1) \, x^{\pm}_{k,m+1,j-1} -\gamma^{\pm} \, m \, x^{\pm}_{k, m, j}, \nonumber
\end{align}
where variables $x^{+}_{k,m,j}$ and $x^{-}_{k,m,j}$  are the fraction of nodes in state $+1$ or $-1$, respectively, with degree $k$, $m$ neighbors in state $-1$ that have been $j$ time steps in the current state. The rates $\beta^{\pm}$ account for the change of state of neighbors ($\pm$) of a node in state $+1$. The rates $\gamma^{\pm}$ are equivalent but for nodes in state $-1$. They can be written as 
\begin{align}
\beta^{+} = & \, \frac{\sum_j \sum_k p_k \sum_{m = 0}^{k} (k - m) \, T^{+}_{k,m,j} \, x^{+}_{k,m,j}}{\sum_j \sum_k p_k \sum_{m = 0}^{k} (k - m) \, x^{+}_{k,m,j}}, \nonumber \\
\beta^{-} = & \, \frac{\sum_j \sum_k p_k \sum_{m = 0}^{k} m \, T^{+}_{k,m,j} \, x^{+}_{k,m,j}}{\sum_j \sum_k p_k \sum_{m = 0}^{k} m \, x^{+}_{k,m,j}}, \nonumber \\
\gamma^{+} = & \, \frac{\sum_j \sum_k p_k \sum_{m = 0}^{k} (k - m) \, T^{-}_{k,m,j} \, x^{-}_{k,m,j}}{\sum_j \sum_k p_k \sum_{m = 0}^{k} (k - m) \, x^{-}_{k,m,j}}, \\
\gamma^{-} = & \, \frac{\sum_j \sum_k p_k \sum_{m = 0}^{k} m \, T^{-}_{k,m,j} \, x^{-}_{k,m,j}}{\sum_j \sum_k p_k \sum_{m = 0}^{k} m \, x^{-}_{k,m,j}}, \nonumber 
\end{align}
 where the degree distribution of the chosen network is $p_k$. The transition rate $T^{\pm}_{k,m,j}$ is for the probability of changing state ($\pm \to \mp$) for an agent of degree $k$, $m$ neighbors in state $-1$ and age $j$, while the aging rate $A^{\pm}_{k,m,j}$ is for the probability of staying in the same state and increasing the internal time ($j \to j + 1$). For the Symmetrical Threshold model, these probabilities do not depend on internal time $j$ (Markovian dynamics):
\begin{align}
    T^{+}_{k,m,j} = & \,  \theta\left(\frac{m}{k} - T\right), \nonumber\\
    T^{-}_{k,m,j} = & \,  \theta\left(\frac{k-m}{k} - T\right) ,\\
    A^{\pm}_{k,m,j} = & \,  1 - T^{\pm}_{k,m,j}. \nonumber
\end{align}
\begin{figure}
    \includegraphics[width=0.8\columnwidth]{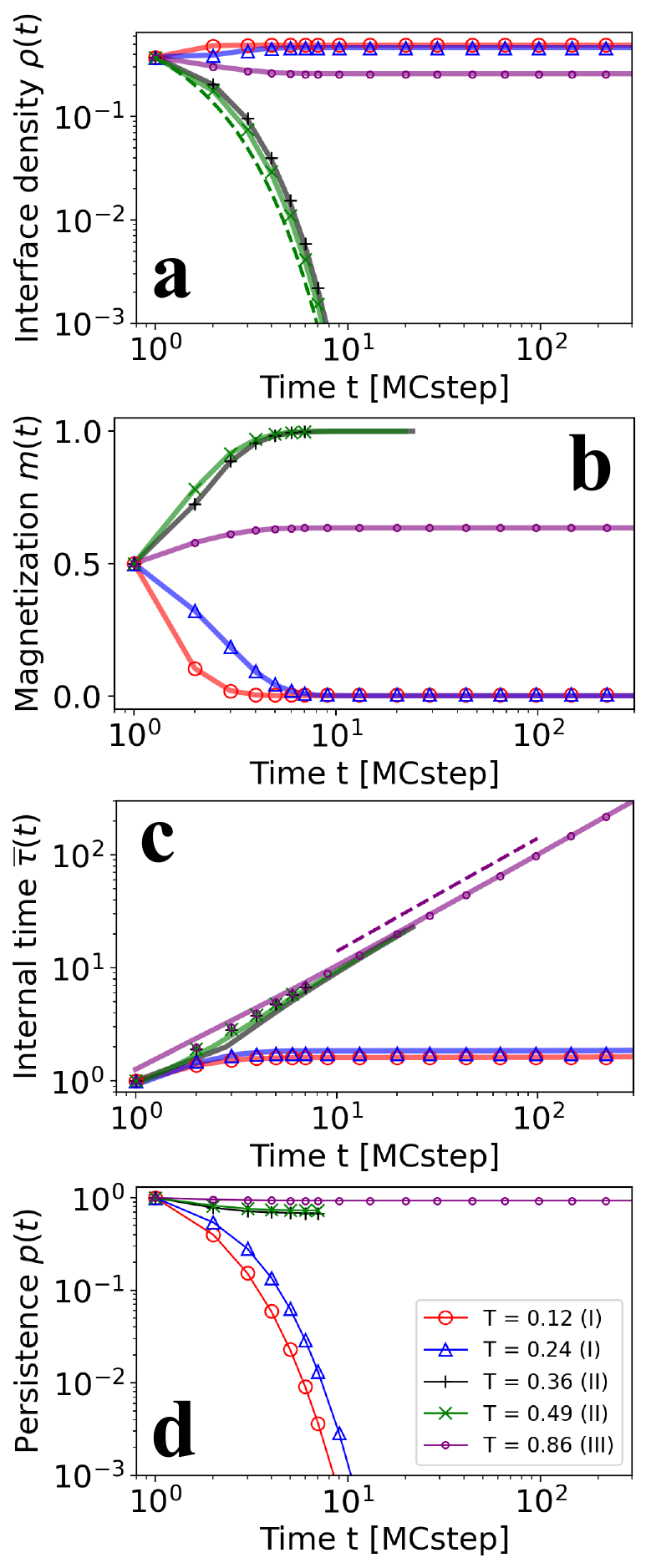}
    \caption{\label{fig:evolution_random}
    Evolution of the average interface density $\rho(t)$ (a), the average magnetization $m(t)$ (b), the mean internal time $\bar{\tau}(t)$ (c), and the persistence $p(t)$ (d) for the Symmetrical Threshold model. The average is computed over $5000$ surviving trajectories (simulations stop when the system reaches the absorbing ordered states). Results for different values of $T$ are plotted with diverse markers and colors: red ($T = 0.12$) and blue ($T = 0.24$) belong to Phase I, green ($T = 0.36$) and grey ($T = 0.49$) belong to Phase II and purple ($T = 0.86$) belongs to Phase III. Solid colored lines are the AME integrated solutions, using Eqs. \eqref{eq:interface}-\eqref{eq:time}. The initial magnetization is $m_0 = 0.5$. The system is on an ER graph with $N = 4 \cdot 10^4$ and mean degree $\langle k \rangle = 8$. The dashed green line in (a) shows $\rho(t) \sim \rho_0 \, e^{-t}$ and the dashed purple line in (c) shows $\bar{\tau}(t) = t$ (purple).}
\end{figure}

If we were not concerned with the internal time dynamics, we can simplify our AME to the one proposed by J. P. Gleeson in Ref. \cite{gleeson-2013} for general binary-state models. Here we keep the internal times for a dynamical characterization of the different phases and as a reference frame for the aging studies in the next section. The primary approximations of this framework are to assume the thermodynamic limit ($N \to \infty$) and  uncorrelated network with negligible levels of clustering. For the complex networks considered, these conditions are satisfied for large $N$, and the differential equations can be solved numerically using standard methods. The mixed order and ordered frozen transitions predicted (solid black lines in Figs. \ref{ER_REG_PD}a and \ref{ER_REG_PD}b, respectively) are in agreement with the numerical simulations. The predicted lines represent the initial and final values of $T$ at which the AME reaches the ordered absorbing states $m_f = \pm 1$. In Fig. \ref{ER_REG_PD}c, we also observe a good agreement between numerically integrated solutions (solid colored lines) and numerical simulations (markers). 

An alternative simpler approximation is to consider a heterogeneous mean-field approximation (HMF) (refer to Appendix \ref{appendix} for further details). Although this approximation captures the qualitative behavior, the numerically integrated solutions do not agree with numerical simulations (see red dashed lines in Figs. \ref{ER_REG_PD}a and \ref{ER_REG_PD}b, and the colored dotted lines in Fig. \ref{ER_REG_PD}c), and the frozen phase is not predicted by this framework. These findings demonstrate that threshold models need approximations beyond mean-field to achieve accuracy, in agreement with the findings in Refs. \cite{gleeson-2007,gleeson-2013,Abella-2022-AME}.

Beyond the stationary states, the previous phases can be characterized by their ordering dynamical properties. To describe the coarsening process, we use the time-dependent average interface density $\rho(t)$ (fraction of links between nodes in different states), the average magnetization $m(t)$, the mean internal time $\bar{\tau}(t)$ (mean time spent in the current state over all the nodes) and the persistence $p(t)$ (fraction of nodes that remain in their initial state at time $t$) \cite{ben-naim-1996}. Fig. \ref{fig:evolution_random} shows the average results obtained from the numerical simulations, starting from an initial magnetization $m_0 = 0.5$. There are 3 regimes with different dynamical properties:
\begin{itemize}
    \item \textbf{Mixed regime (Phase I):} It corresponds to Phase I in the static phase diagram and it is characterized by fast disordering dynamics, which is reflected by an exponential decay of the persistence. The interface density, the magnetization, and the mean internal time exhibit fast dynamics towards their asymptotic values in the dynamically active stationary state (see $T = 0.12, 0.24$ in Fig. \ref{fig:evolution_random});
    \item \textbf{Ordered regime (Phase II):} It coincides with Phase II in the static diagram and it is characterized by an exponential decay of the interface density. The magnetization tends to the ordered absorbing state based on the initial majority, and the mean internal time tends to scale as $\bar{\tau}(t) \sim t$. Persistence in this phase decays until a plateau that corresponds to the initial majority that reaches consensus (since this fraction of nodes does not change state from the initial condition). When consensus is reached, the surviving trajectory is stopped (see $T = 0.36, 0.49$ in Fig. \ref{fig:evolution_random});
    \item \textbf{Frozen regime (Phase III):} This regime corresponds to Phase III and it is characterized by an initial ordering process followed by the stop of the dynamics, with constant values of the metrics. The only exceptions are the mean internal time that grows as $\bar{\tau}(t) \sim t$ (see $T = 0.86$ in Fig. \ref{fig:evolution_random}) and the persistence.
\end{itemize}
Using the numerically integrated solutions of AME ($x^{\pm}_{k,m,j}(t)$), we can compute the magnetization $m(t)$, the interface density $\rho(t)$, and the mean internal time $\bar{\tau}$:
\begin{align}
    \rho(t) =  & \,  \frac{2 \, \sum_j \sum_k p_k \sum_m  m x^{+}_{k,m,j}}{ \sum_j \sum_k p_k \sum_m  k (x^{+}_{k,m,j} + x^{-}_{k,m,j})},\label{eq:interface}\\
    \nonumber\\
    m(t) = & \,  2 \sum_j \sum_k p_k \sum_m x^{+}_{k,m,j} - 1 \label{eq:magne}\\
    \nonumber\\
    \bar{\tau} (t) = & \,  \sum_j \sum_k p_k \sum_m j \left(x^{+}_{k,m,j} + x^{-}_{k,m,j}\right).\label{eq:time}
\end{align}
All metrics exhibit a strong agreement between the numerical simulations and the integrated solutions (see solid lines in Fig. \ref{fig:evolution_random}). However, the persistence cannot be directly calculated from the integrated solutions. This is because the fraction of persistent nodes at time $t$ corresponds to the fraction of nodes with internal time $j = t$, which is at an extreme of the age distribution at each time step, since $x^{\pm}_{k,m,j}(t) = 0$ for $j > t$. Therefore, the computation of this measure requires a more sophisticated analysis using extreme value theory \cite{haan2006extreme}.

We note that the dynamical characterization discussed above holds for all possible $m_0$ except for the symmetric initial condition $m_0 = 0$. In this case, an order-disorder transition arises at a critical mean degree $k_c$, whose value depends on the size of the system $N$ \cite{Konstantin}. 

\section{\label{Symmetrical Threshold model with aging} Symmetrical Threshold model with aging}

Aging refers to the property of agents becoming less likely to change their state the longer they have remained in that state \cite{stark-2008,artime-2017,artime-2018,peralta-2020A,peralta-2020C,chen-2020,Abella-2022,Abella-2022-AME}. In contrast to the original model, which assumes that agents update their state at a constant rate, this model introduces an activation probability $p_A (j)$ that is inversely proportional to the agent's internal time $j$. At each time step, the following two steps are performed:
\begin{enumerate}
    \item A node $i$ with age $j$ is selected at random and activated with probability $p_A(j)$;
    \item If the fraction of neighbors in a different state is greater than the threshold $T$, the activated node changes its state from $s_i$ to $-s_i$ and resets its internal time to $j=0$.
\end{enumerate}
We set the activation probability to $p_A(j) = 1/(j+2)$ with the aim of recovering a fat-tailed inter-event time distribution, as observed in simple contagion models \cite{fernandez-gracia-2011,artime-2017}. 

\subsection{Mean-field}

\begin{figure}[ht]
\includegraphics[width=\columnwidth]{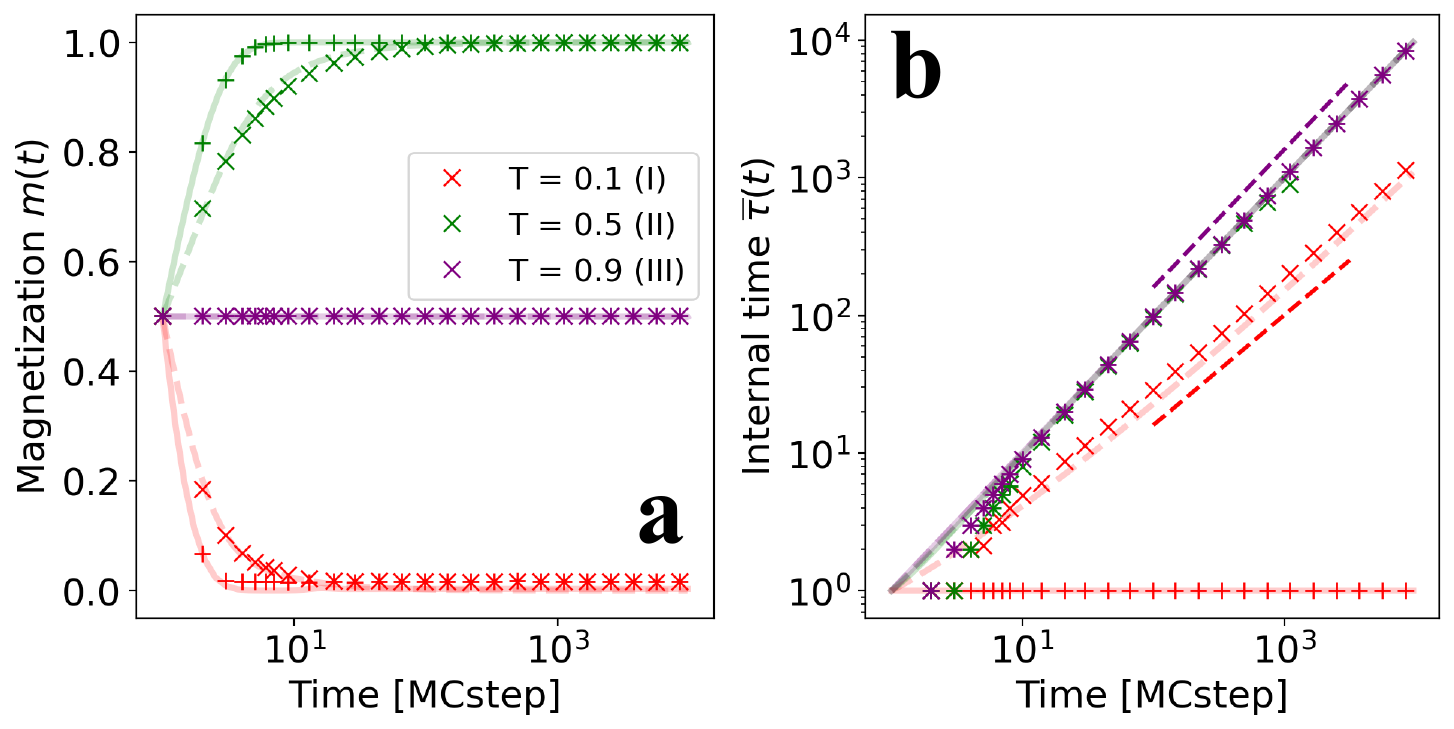}
\caption{\label{fig:COM_AGING} Evolution of the average magnetization $m(t)$ (a) and the mean internal time $\bar{\tau}(t)$ (b) in a complete graph of $N=2500$ nodes. Results are shown for the Symmetrical Threshold Model (pluses) and the version with aging (crosses) obtained from simulations. Different colors correspond to different values of the threshold $T$: red ($T = 0.1$) belongs to Phase I, green ($T = 0.5$) belongs to Phase II, and purple ($T = 0.9$) to Phase III. The initial magnetization is fixed at $m_0 = 0.5$. The solid and dashed lines correspond to the numerically integrated solutions from Eq. \ref{eq:HMFaging2} for the original model ($p_A(j) = 1$) and the version with aging ($p_A(j) = 1/(t+2)$), respectively.}
\end{figure}

Figure \ref{fig:COM_AGING} compares the evolution of the average magnetization and mean internal times on a complete graph of the original Symmetrical Threshold model and the version with aging in phases I, II and III. We observe that, for all considered threshold values, aging introduces a delay. However, the final stationary state coincides with the one observed for the original model. To explain these dynamics, we use a heterogeneous mean-field approach that considers the effects of aging (HMFA), as in Ref. \cite{chen-2020} for other binary-state models (we use a general HMF description to be applied for a complete graph and to random networks in next section). In this case, the AME does not work well due to the high density of the network. For a general network with degree distribution $p_k$, we define the fraction of agents in state $\pm 1$ with $k$ neighbors and age $j$ at time $t$ as $x^{\pm}_{k,j} (t)$. The probability of finding a neighbor in state $\pm 1$ is $\tilde{x}^{\pm}$, which can be written as 
\begin{equation}
    \tilde{x}^{\pm} = \sum_k p_k \frac{k}{\langle k \rangle} \,  \sum_{j=0}^{\infty} x^{\pm}_{k,j},
\end{equation}
where $\langle k \rangle$ is the mean degree of the network. The transition rate $\omega_{k,j}^{\pm}$ for a node with state $\pm 1$, degree $k$ and age $j$ to change state is given by 
\begin{equation}
    \omega_{k,j}^{\pm} = p_{A} (j) \,  \sum_{m=0}^{k} \theta\left(\frac{m}{k} - T\right) \,  B_{k,m}[\tilde{x}^{\mp}],
\end{equation}
where $B_{k,m}[x]$ is the binomial distribution with $k$ attempts, $m$ successes, and with the probability of success $x$. In our model, there are two possible events for a node with degree $k$ and age $j$:
\begin{itemize}
    \item It changes state and the age is reset to $j = 0$;
    \item It remains at its state and the age increases by one time step $j = j + 1$.
\end{itemize}
According to these possible events, we can write the rate equations for the variables  $x^{\pm}_{k,j}$ and $x^{\pm}_{k,0}$ as
\begin{align}
\label{eq:HMFaging2}
    \frac{dx^{\pm}_{k,0}}{dt} & = \sum_{j=0}^{\infty} x^{\mp}_{k,j} \,  \omega_{k,j}^{\mp} - x^{\pm}_{k,0} \qquad \mbox {for zero age},\\
    \frac{dx^{\pm}_{k,j}}{dt} & =  x^{\pm}_{k,j-1} \,  ( 1 - \omega_{k,j-1}^{\pm}) - x^{\pm}_{k,j} \qquad \mbox {if } j > 0. \nonumber 
\end{align}
It can be shown from Eq. \eqref{eq:HMFaging2} that the stationary solution  for the fraction of agents in state $+1$, $x_f$, obeys the following implicit equation for a complete graph (see Appendix \ref{appendix_HMFA} for a detailed explanation):
\begin{equation}
    x_f = \frac{F(1 - x_f)}{F(x_f) + F(1-x_f)},
    \label{eq:x_f}
\end{equation}
where,
\begin{equation}
    F(x) = 1 + \sum_{j=1}^{\infty} \prod_{a=0}^{j-1} \left( 1 - p_A(a) \, \sum_{m = (N-1)T}^{N-1} B_{N-1,m}[x] \right).
    \label{eq:F(A)}
\end{equation}

A solution of Eq. \eqref{eq:x_f} can be obtained numerically using standard methods, as in Ref. \cite{chen-2020}. The final magnetization is calculated as $m_f = 2 \,x_f - 1$. With this method, we obtain that the phase diagram for the model with aging is the same as for the original model (refer to Fig. \ref{COM_LAT_PD}a). As a technical point, we note that a truncation of the summation of the variable $j$ in Eq. \eqref{eq:F(A)} is required for the numerical resolution of the implicit equation. The higher the maximum age considered $j_{\rm max}$, the higher the accuracy. With $j_{\rm max} = 5 \cdot 10^4$, the transition lines predicted by this mean-field approach show great accuracy. Moreover, by numerically integrating Eqs. \eqref{eq:HMFaging2}, the dynamical evolution of the magnetization and mean internal time can be obtained. Fig. \ref{fig:COM_AGING} shows the agreement between integrated solutions and Monte Carlo simulations of the system both for the aging and non-aging versions. It should be noted that, while aging introduces only a dynamical delay for the magnetization $m(t)$, the mean internal time $\bar{\tau}(t)$ in Phase I shows a different dynamical behavior with aging than in the original model. In this phase, due to the low value of $T$, the agents selected randomly will change their state (as they fulfill the threshold condition) and reset their internal time. Consequently, while the internal time fluctuates around a stationary value for the original model, when aging is incorporated, due to the activation probability $p_A(j)$ chosen, the mean internal time increases following a recursive relation (Eq. \eqref{eq:RR}). We refer to Appendix \ref{appendix_RR} for a derivation of this result.

\subsection{\label{sec:Complex networks aging} Random networks}

\begin{figure}
     \includegraphics[width=0.7\columnwidth]{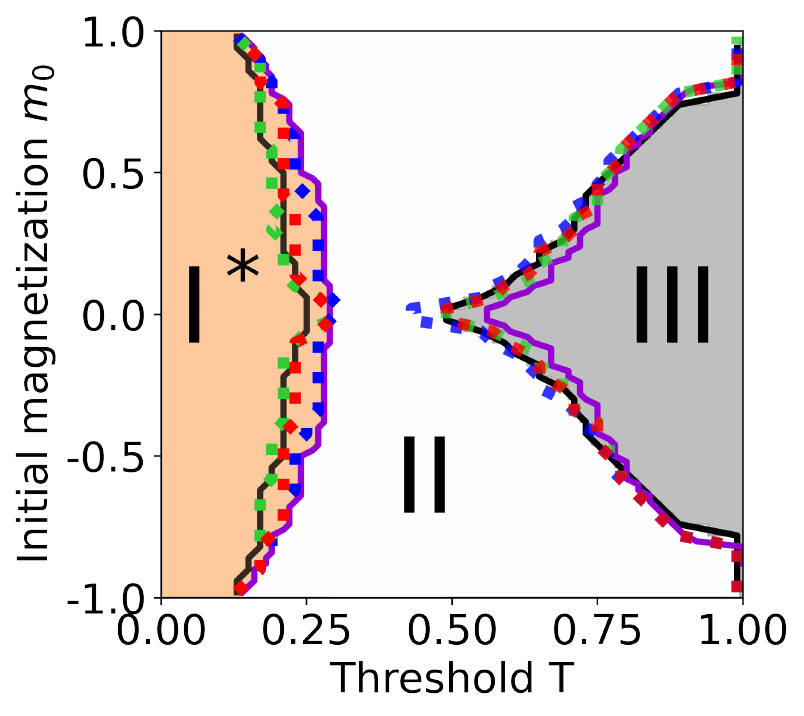}
     \caption{\label{ER_REG_PDAGING} Phase diagram of the Symmetrical Threshold with aging model in an ER graph of $N = 4 \cdot 10^4$ nodes and $\langle k \rangle = 8$. The blue, red, and green dotted lines show the borders of Phase II (first and last value of $T$ where the system reaches the absorbing ordered state for each $m_0$) computed from numerical simulations evolving until $t_{\rm max} = 10^3$, $10^4$ and $10^5$ time steps, respectively. Black solid lines show AME solution integrated $10^5$ time steps. Phase ${\rm I}^{*}$, II and III correspond with the orange, white and gray areas, respectively. The solid purple lines are the mixed-ordered and ordered-frozen critical lines for the non-aging version of the model.}
\end{figure}

In contrast to the results obtained in a complete graph, aging effects have a significant impact on the phase diagram of the model on random networks. In Fig. \ref{ER_REG_PDAGING}, we show the borders of Phase II (first and last value of $T$ where the system reaches the absorbing ordered state for each $m_0$) obtained from Monte Carlo simulations running up to a maximum time $t_{\rm max}$ (dotted colored lines). Reaching the stationary state in this model requires a large number of steps and it has a high computational cost. The two borders of Phase II exhibit different behavior as we increase the time cutoff $t_{\rm max}$: while the ordered-frozen border does not change with different $t_{\rm max}$, the mixed-ordered border is shifted to lower values of $T$ as we increase the time cutoff $t_{\rm max}$. Our results suggest that Phase I is actually replaced in a good part of the phase diagram by an ordered phase in which the absorbing state $m_f = \pm 1$ is reached after a large number of time steps. The ordered-frozen border is now slightly shifted to lower values of the threshold $T$ due to aging. Similar results are found for a RR graph (see Appendix \ref{RR_phase_diagram}). The dependence of the results with $t_{\rm max}$ calls for a characterization of different phases in terms of dynamical properties rather than by the asymptotic value of the magnetization.

\begin{figure}
     \centering
     \includegraphics[width=0.75\columnwidth]{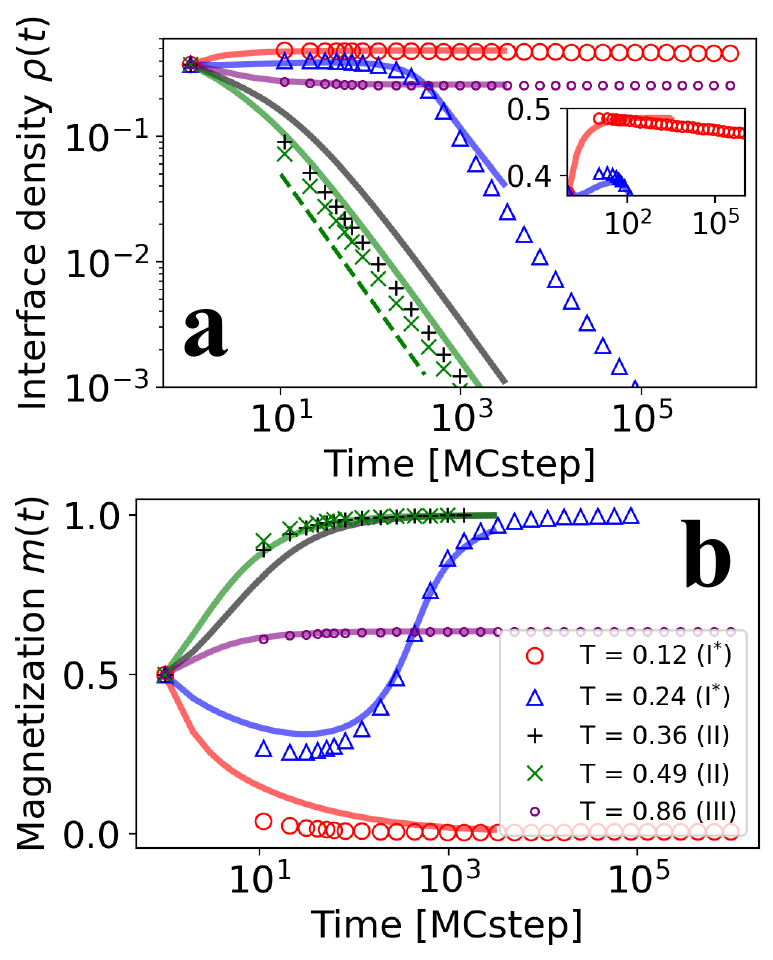}
     \caption{\label{fig:evolution_random_aging} Evolution of the average interface density $\rho(t)$ (a) and the average magnetization $m(t)$ (b) for the Symmetrical Threshold model with aging. The average is computed over $5000$ surviving trajectories (simulations stop when the system reaches the absorbing ordered states) for different values of $T$, shown by different markers and colors: red ($T = 0.12$) and blue ($T = 0.24$) belong to Phase ${\rm I}^{*}$, green ($T = 0.36$) and grey ($T = 0.49$) belong to Phase II and purple ($T = 0.86$) belong to Phase III. The inset in (a) shows a close look to the evolution for $T = 0.12$, in linear-log scale. Solid colored lines are the AME integrated solutions for $10^4$ time steps, using Eqs. \ref{eq:interface} - \ref{eq:magne}. The initial magnetization is $m_0 = 0.5$. The system is on an ER graph with $N = 4 \cdot 10^4$ and mean degree $\langle k \rangle = 8$. The dashed green line in (a) shows $\rho(t) \sim \rho_0 \, t^{-1}$.}
\end{figure}

\begin{figure}
    \includegraphics[width=0.8\columnwidth]{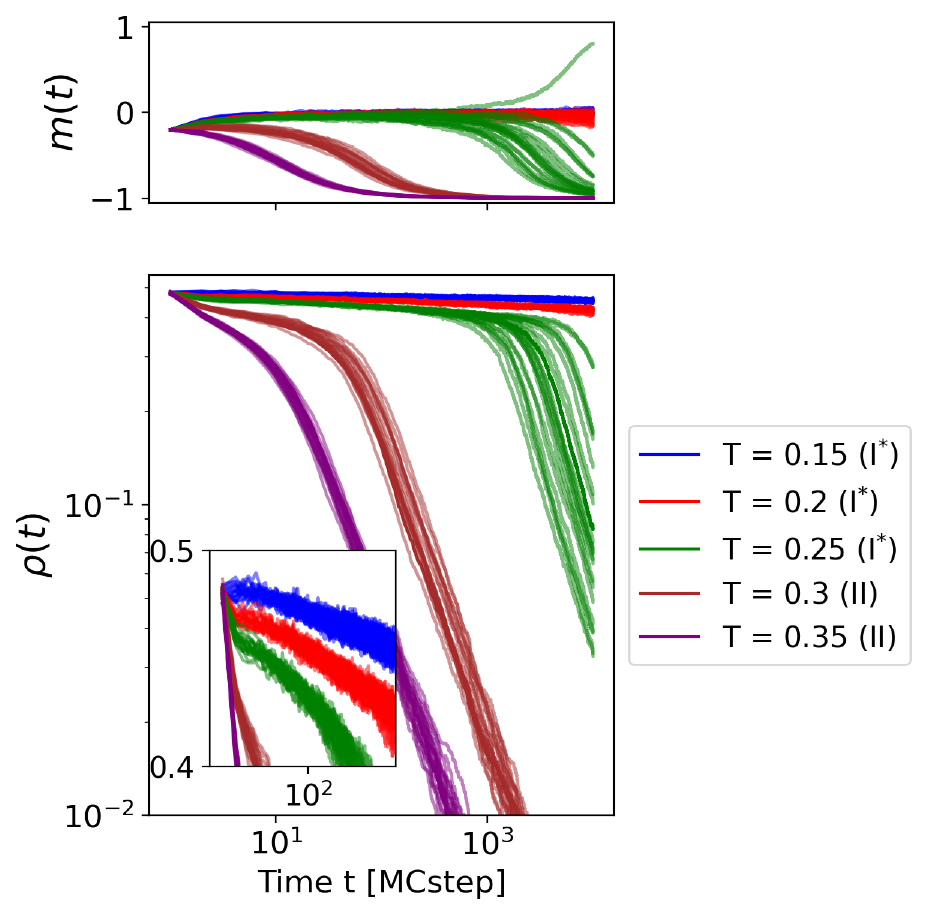}
    \caption{\label{fig:mixed_phase} Interface density $\rho(t)$ (lower) and magnetization $m(t)$ (upper) trajectories for different values of the threshold $T$ ($m_0 = -0.2$) using the Symmetrical Threshold model with aging. Different colors indicate different values of $T$. The inset shows a close look at the logarithmic-like decay, shown in linear-log scale. The system is an ER graph with $N = 4 \cdot 10^4$ and mean degree $\langle k \rangle = 8$.}
\end{figure}

Figure \ref{fig:evolution_random_aging} shows the time evolution of our ordering metrics. The dynamical properties are largely affected by the aging mechanism. In terms of the evolution, we find the following regimes:
\begin{itemize}
    \item \textbf{Initial mixing regime (Phase ${\rm {\bf I}}^{*}$):} It is characterized by two dynamical transient regimes: a fast initial disordering dynamics followed by a slow ordering process. After the initial fast disordering stage, the average interface density exhibits a very slow (logarithmic-like) decay. Later, due to the finite size of the system the average interface density follows a power law decay with time, where $\rho(t)$ scales as $t^{-1}$. This phase exists for the same domain of parameters ($m_0$, $T$) as Phase I (orange region in Fig. \ref{ER_REG_PDAGING}) in the model without aging (see $T = 0.12, 0.24$ in Fig. \ref{fig:evolution_random_aging});
    \item \textbf{Ordered regime (Phase II):} It is characterized by a power-law interface decay, where $\rho(t)$ scales as $t^{-1}$. The magnetization tends to the ordered absorbing state according to the initial majority (see $T = 0.36, 0.49$ in Fig. \ref{fig:evolution_random_aging});
    \item \textbf{Frozen regime (Phase III):} It is characterized by an initial tendency towards the majority consensus, but very fast reaches an absorbing frozen configuration (see $T = 0.86$ in Fig. \ref{fig:evolution_random_aging}).
\end{itemize}
The main effect of aging is that the mixed states of Phase I are no longer present, at least not for the type of networks that we are analyzing here. We will show later that Phase I reemerges in denser graphs. Instead, for sparse graphs, we observe a new Phase ${\rm I}^{*}$ in which the system initially disorders and later orders until reaching the absorbing states $m_f = \pm 1$. This behavior is shown in Fig. \ref{fig:evolution_random_aging} for $T = 0.12$ and $0.26$. For $T = 0.12$, the system initially disorders, and then the interface density follows a logarithmic-like decay (see inset in Fig. \ref{fig:evolution_random_aging}a). Due to the slow decay, the system stays in this transient regime even after $10^{6}$ time steps, and the fall to the absorbing states is not seen. Similarly, for $T = 0.26$ the disordering process stops and then the system gradually evolves towards a fully ordered state. For this value of $T$, the logarithmic-like decay is not appreciated and we just observe the power-law decay due to the finite size of the system. The difference between $T = 0.12$ and $T = 0.26$ comes from the fact that in this Phase ${\rm I}^{*}$, the decay of $\rho$ becomes faster as we increase the threshold $T$ (see inset in Fig. \ref{fig:mixed_phase}). Notice the different interface decay in Fig. \ref{fig:mixed_phase} between values of $T < 0.3$ (Phase ${\rm I}^{*}$), where all trajectories show a logarithmic-like decay of $\rho(t)$ in a transient regime, and $T \geq 0.3$ (Phase II), where trajectories from the initial condition exhibit ordering dynamics towards the consensus of the majority. Moreover, we observe that in Phase ${\rm I}^{*}$, the initial magnetization $m_0$ introduces a bias to the stochastic process, implying that the larger $m_0$ in absolute value, the larger the number of realizations that reach the absorbing state with the same sign of $m_0$. However, the system can still reach the absorbing state of the opposite sign of $m_0$ (initial minority), as shown in the trajectories with $T = 0.25$ in Fig. \ref{fig:mixed_phase}.
In Phase II, the system asymptotically orders for any initial condition as in the original model, but the dynamical properties are modified due to the presence of aging: the exponential decay of the interface density is replaced by a slow power-law decay, where the exponents of the exponential and the power-law are found to be the same. Contrary, the dynamical properties of Phase III are not affected by the presence of aging. The temporal magnitudes analysis (mean internal time and persistence) can be found in Appendix \ref{sec:temporal_dynamics}.

To account for the results of our Monte Carlo simulations, we use the same mathematical framework as described in Equation \eqref{eq:AME_age}. According to the update rules of the Symmetrical Threshold Model with aging, the transition probabilities now depend on the age $j$, as given by the activation probability  $p_A (j)$:
\begin{align}
    T^{+}_{k,m,j} = & \, p_A(j) \, \theta(m/k - T) ,\nonumber\\
    T^{-}_{k,m,j} = & \, p_A(j) \, \theta((k-m)/k - T) ,\\
    A^{\pm}_{k,m,j} = & \,  1 - T^{\pm}_{k,m,j}. \nonumber
\end{align}
We show in Figure \ref{ER_REG_PDAGING} the mixed-ordered and ordered-frozen transition lines predicted by the integration of the AME equations until a time cutoff $t_{\rm max}$. We find good agreement between the theoretical predictions and the simulations both for ER and RR networks (see RR results in Appendix \ref{RR_phase_diagram}). Regarding dynamical properties, the AME integrated solutions exhibit a remarkable concordance with the evolution of all the metrics as shown in Figure \ref{fig:evolution_random_aging}. Minor discrepancies between the numerical simulations and the integrated solutions can be attributed to the assumption of an infinitely sized system in the AME.
\begin{figure}[b]
     \includegraphics[width=\columnwidth]{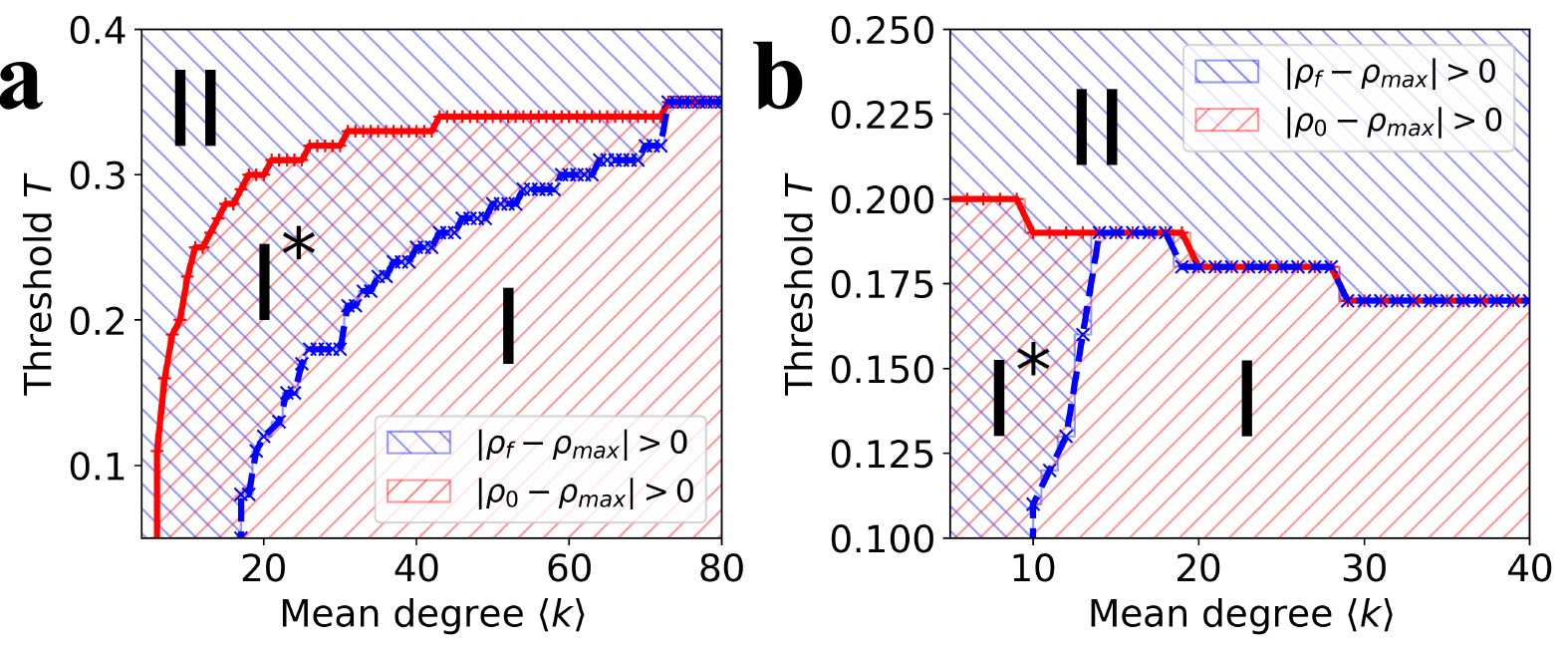}
     \caption{\label{fig:PD_Z}  Critical threshold $T_c$ dependence  with the mean degree $\langle k \rangle$ for the Symmetrical Threshold model with aging for an ER graph with $N = 4 \times 10^4$ nodes for an initial magnetization of $m_0 = 0.25$ (a) and $m_0 = 0.75$ (b). The blue and red markers indicate the borders of phases I and II, which coincide for a sufficiently large value of the mean degree. The hatched area corresponds to the fulfillment of the inequality in the legend.}
\end{figure}

The numerical results discussed so far are for random networks with average degree $\langle k \rangle = 8$. According to them and to the analytical insights, one can conclude that aging significantly changes the phase diagram for sparse networks. However, we know that the mean-field (fully connected) model with aging shows the same phase diagram as the model without aging. This implies that, for ER graphs, as the mean degree $\langle k \rangle$ approaches $N$, Phase ${\rm I}^{*}$ must disappear. Therefore, the combined effects of increasing the mean degree and introducing aging need to be investigated in more detail. Phase II is distinguishable from phases I and ${\rm I}^{*}$ because the system initially orders, i.e., $|\rho_0 - \rho_{\rm max}| = 0$, where $\rho_{\rm max}$ is the maximum value attained by the interface density during the dynamical evolution. In contrast, Phase I is distinguished from Phases ${\rm I}^{*}$ and II because the system remains disordered, i.e., $|\rho_{\rm max} - \rho(t_{\rm max})| \approx 0$. Thus, Phase ${\rm I}^{*}$ is the only phase among these three where $|\rho_0 - \rho_{\rm max}| > 0$ and $|\rho_{\rm max} - \rho(t_{\rm max})| > 0$. Using this criterion, we studied the dependence of the critical threshold  $T_c$ on the mean network degree defining the transition lines between phases I, ${\rm I}^{*}$, and II (see Fig. \ref{fig:PD_Z}). In the absence of aging, the red line in Fig. \ref{fig:PD_Z} gives the value of the mixed-ordered transition line $T_c$. When aging is included, at low degree values, Phase I is replaced by ${\rm I}^{*}$, as expected. However, as the mean degree increases, Phase I emerges despite the presence of aging, leading to the coexistence of phases I and ${\rm I}^{*}$ in the same phase diagram over a range of mean degree values. As the mean degree is further increased, a critical value is reached where Phase ${\rm I}^{*}$ is no longer present, and the discontinuous transition I-II occurs at the same value than in the model without aging. Importantly, this critical mean degree at which Phase ${\rm I}^{*}$ disappears, depends significantly on the initial magnetization $m_0$.

\section{\label{sec: Symmetrical Threshold model in a Moore_Lattice}  Symmetrical Threshold model in a Moore Lattice}

\begin{figure}[b]
     \includegraphics[width=\columnwidth]{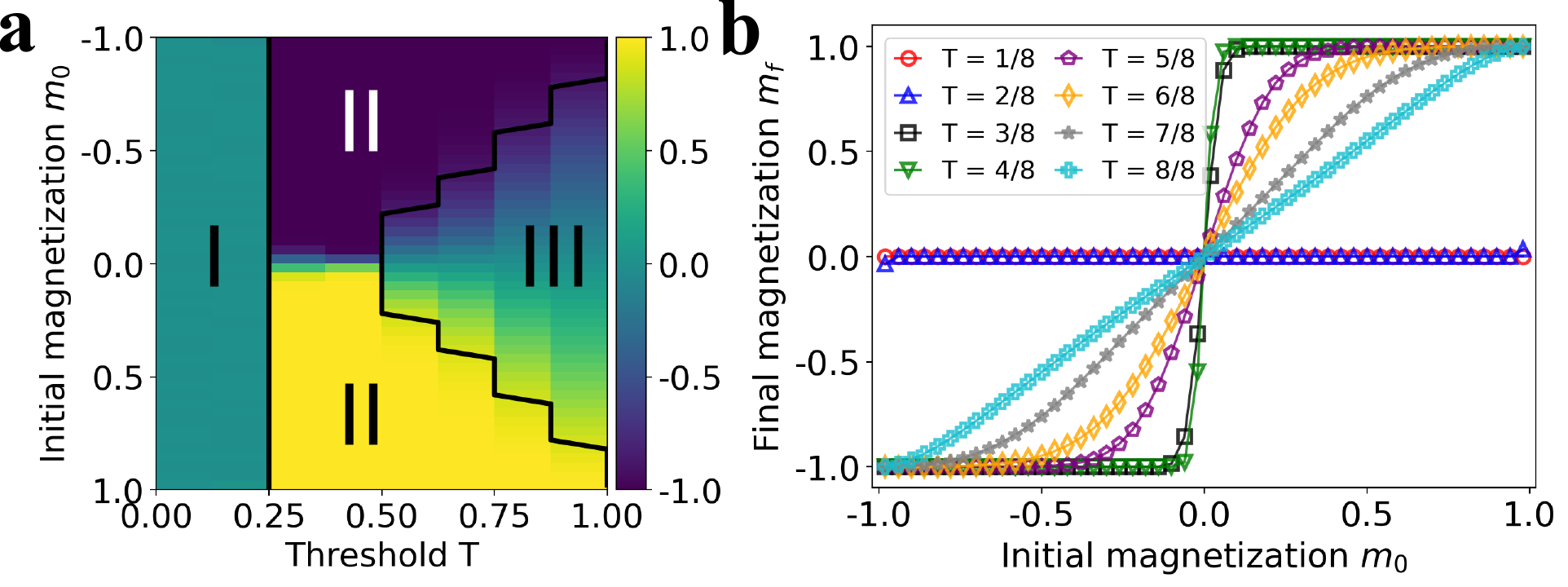}
     \caption{\label{LAT_PD} (a) Phase diagram of the Symmetrical Threshold model in a Moore lattice of size $N = L \times L$, with $L = 100$. The color map indicates the value of the average final magnetization $m_f$. Solid black lines are the borders of Phase II (first and last value of $T$ where the system reaches the absorbing ordered state for each $m_0$), computed from the numerical simulations. (b) Average final magnetization $m_f$ as a function of the initial magnetization $m_0$ for the discrete values of the threshold $T$ (indicated with different colors and markers) in a Moore lattice of the same size. Average performed over 5000 realizations.}
\end{figure}

We consider next the Symmetrical Threshold model in a Moore lattice, which is a regular 2-dimensional lattice with interactions among nearest and next-nearest neighbors ($k=8$).  From numerical simulations, we obtain a phase diagram (Fig. \ref{LAT_PD}a) that is consistent with our previous results in random networks. The system undergoes a mixed-ordered transition at a threshold value $T_{c} = 2/8$  which is independent of the value of the initial magnetization $m_0$. When $T > 4/8$, the system undergoes an ordered-frozen transition at a critical threshold $T_{c}^{*}$, which depends on $m_0$ (similarly to what happens in random networks). The final magnetization $m_f(m_0)$ (Fig. \ref{LAT_PD}b) also shows a dependence on $m_0$ similar to the one found in RR networks (Fig. \ref{ER_REG_PD}c).

\subsection{Original model without aging}

\begin{figure*}
     \centering
     \includegraphics[width=0.85\linewidth]{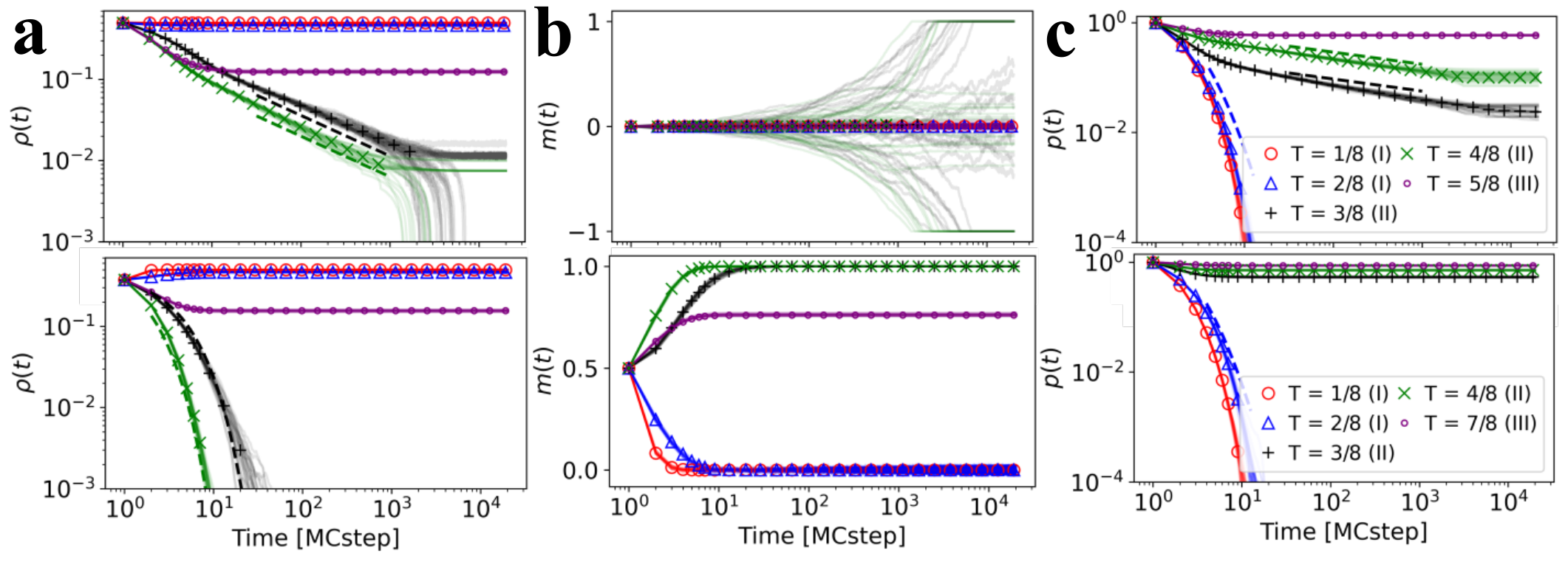}
     \caption{\label{fig:evolution_lattice} Evolution of the interface density $\rho(t)$ (a), the average magnetization $m(t)$ (b) and the persistence $p(t)$ (c) for the Symmetrical Threshold model in a Moore lattice starting from a random configuration with $m_0 = 0$ (upper) and $m_0 = 0.5$(lower). We plot 50 different trajectories in solid lines and the average of $5000$ surviving trajectories (simulations stop when the system reaches the absorbing ordered states) in different markers. Different colors and markers indicate different threshold values: red ($T = 1/8$) and blue ($T = 2/8)$ belong to Phase I, green ($T = 3/8$) and black ($T=4/8$), and purple ($T = 5/8, 7/8$) belong to Phase III. The average magnetization $m(t)$ is computed according to the two symmetric absorbing states. System size is fixed at $N = L \times L$, $L = 200$. The dashed lines in (a-upper) are $\rho(t) \sim at^{-1/2}$ with $a = 0.36$ (black) and $a = 0.2$ (green), in (a-lower) are $\rho \sim \exp(-\alpha \cdot t)$ with $\alpha = 0.5$ (black) and $\alpha = 0.8$ (green), (c) are $p(t) \sim \exp(- \ln(t)^2)$ (blue).}
\end{figure*}

Fig. \ref{fig:evolution_lattice} shows the results from numerical simulations (for $m_0 = 0$ and $0.5$) for the average interface density, the magnetization, and the persistence (the internal time shows the same results as in random graphs). Dynamical properties change significantly for different values of the threshold and initial magnetization $m_0$. Similarly to the case of random networks, we find three different regimes corresponding to the three phases, but with some properties different from the results on  random networks:
\begin{itemize}
    \item \textbf{Mixed regime (Phase I):} It is characterized by fast disordering dynamics with a persistence decay $p(t) \sim \exp(- \ln(t)^2)$. The interface density and the magnetization exhibit fast dynamics towards their asymptotic values in the dynamically active stationary state (see $T = 1/8,2/8$ in Fig. \ref{fig:evolution_lattice});
    \item \textbf{Ordered regime (Phase II):} It is characterized by an exponential or power-law decay of the interface density, depending on the initial condition. The magnetization tends to the absorbing ordered state (see $T = 3/8,4/8$ in Fig. \ref{fig:evolution_lattice});
    \item \textbf{Frozen regime (Phase III):} It is characterized by an initial ordering process, but the system freezes fast (see $T = 5/8$ in Fig. \ref{fig:evolution_lattice}).
\end{itemize}

In particular, in Phase II the persistence and interface density decay as a power law, $p(t) \sim t^{-0.22}$ and $\rho(t) \sim t^{-1/2}$ for $m_0 = 0$ (as in Refs. \cite{stauffer-1994,derrida-1995A,derrida-1995B,derrida-1997}). For a biased initial condition ($m_0 = 0.5$), $p(t)$ decays to the initial majority fraction (which corresponds to the state reaching consensus) and $\rho(t)$ follows an exponential decay. Note that, for $m_0 = 0$, not all trajectories reach the ordered absorbing states ($m_f=\pm 1$). There exist other absorbing configurations as, for example,  a flat interface configuration for $T = 4/8$, no agent will be able to change, and the system remains trapped in this state. This result is not observed for $m_0 > 0$.

Contrary, phases I and III show similar dynamics for a symmetrical ($m_0 = 0$) and biased ($m_0 = 0.5$) initial conditions. In Phase I, the system shows disordering dynamics with a persistence decay similar to the one exhibited for the Voter model in a lattice \cite{ben-naim-1996} while in Phase III, the system exhibited freezing dynamics with an initial tendency towards the majority consensus.

Due to the lattice structure and high clustering, the mathematical tools used in previous sections for random networks cannot be applied in the case  of a regular lattice. In this case, we restrict ourselves to the results of numerical simulations.


\subsection{The role of aging}

\begin{figure}
     \includegraphics[width=\columnwidth]{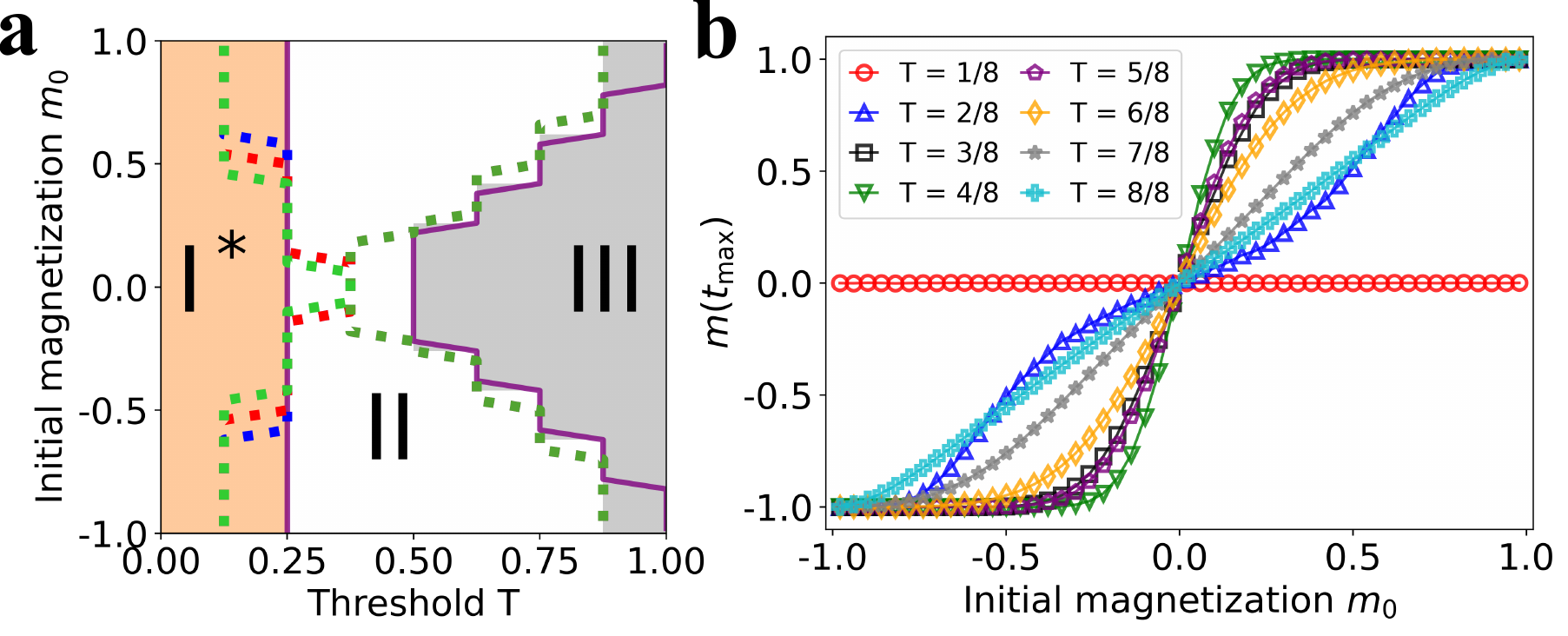}
     \caption{\label{LAT_PDAGING} (a) Phase diagram of the Symmetrical Threshold model with aging in a Moore lattice of $N = L \times L$, with $L = 100$. The blue, red and green dotted lines show the borders of Phase II (first and last value of $T$ where the system reaches the absorbing ordered state for each $m_0$) from numerical simulations evolving until $t_{\rm max} = 10^3$, $10^4$ and $10^5$ time steps, respectively. Phase ${\rm I}^{*}$, II and III correspond with the orange, white and gray areas, respectively. The solid purple lines are the mixed-ordered and ordered-frozen critical lines for the Symmetrical threshold model (from Fig. \ref{LAT_PD}) (b) Average magnetization at time $t_{\rm max}$ ($m_f(t_{\rm max})$) as a function of the initial magnetization $m_0$ for different values of the threshold $T$ (indicated with different colors and markers) in a Moore lattice of $N = L \times L$, with $L = 100$. The numerical simulations are obtained until $t_{\rm max} = 10^4$ MC steps. Average performed over 5000 realizations.}
\end{figure}

\begin{figure*}
     \centering
     \includegraphics[width=0.85\linewidth]{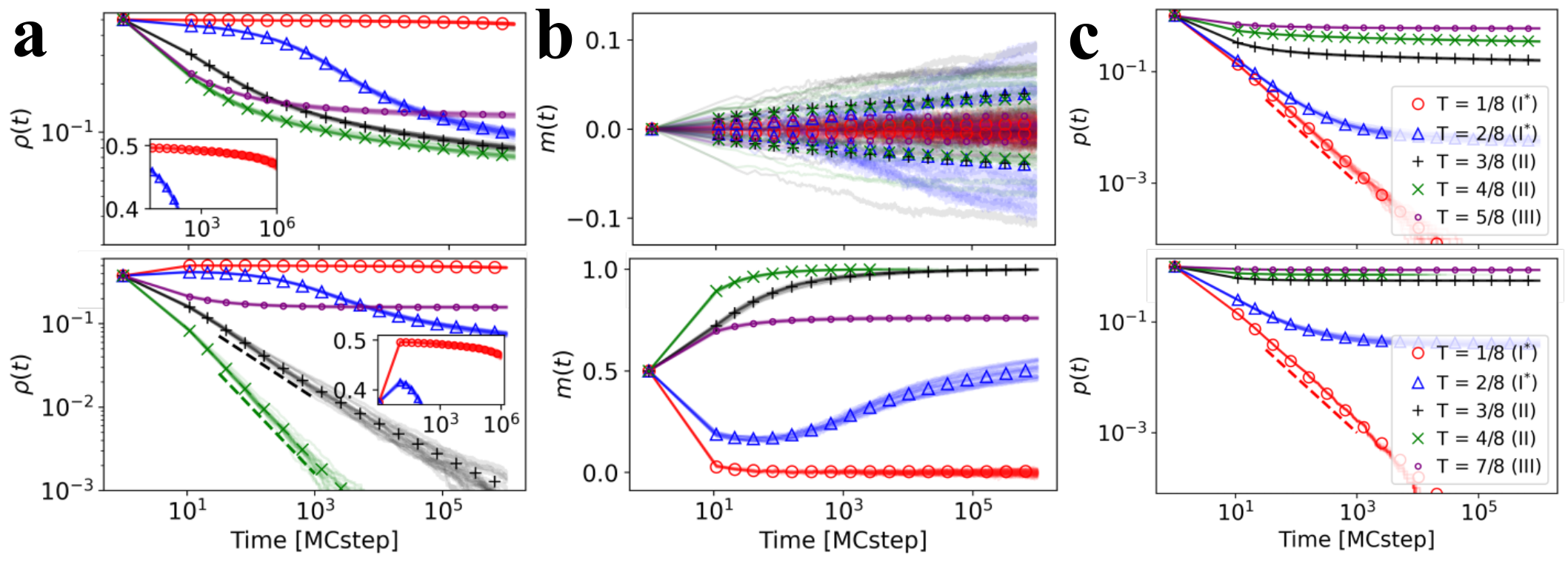}
     \caption{\label{fig:evolution_lattice_aging} Evolution of the average interface density $\rho(t)$ (a), the average magnetization $m(t)$ (b) and the persistence $p(t)$ (c) for the Symmetrical model with aging in a Moore lattice starting from a random configuration with $m_0 = 0$ (upper) and $m_0 = 0.5$(lower). We plot 30 different trajectories in solid lines and the average over $5000$ surviving trajectories in symbols. Colors and symbols indicate different threshold values: red ($T = 1/8$) and blue ($T = 2/8)$ belong to Phase ${\rm I}^{*}$, green ($T = 3/8$), and black ($T=4/8$) belong to Phase II, and purple ($T = 5/8, 7/8$) belong to Phase III. The average magnetization is computed according to the two symmetric absorbing states. The insets in (a) show a close look at the evolution for $T = 0.12$, in linear-log scale. System size is fixed at $N = L \times L$, $L = 200$. The dashed lines in (a) are $\rho \sim t^{-\alpha}$ with $\alpha = 0.5$ (black) and $\alpha = 0.8$ (green), and in (c) are $p(t) \sim t^{-1}$ (red). Simulations stop when the system reaches the absorbing ordered states.}
\end{figure*}

We show in Figure \ref{LAT_PDAGING}a borders of Phase II obtained from numerical simulations running up to a time $t_{\rm max}$ (dotted colored lines). Similarly to the behavior observed in random networks, the mixed-ordered border is shifted to lower values of $T$ as we increase the simulation time cutoff $t_{\rm max}$. Thus, Phase I is replaced by an ordered phase due to the aging mechanism. Examining the dependence of the final value of the magnetization on its initial condition  $m_f(m_0)$  (Figure \ref{LAT_PDAGING}b), one can conclude that the mixed phase is still present, at least transiently, as in the initially disordering phase described in the previous section (Phase ${\rm I}^{*}$). Phase II is again characterized by an asymptotically ordered state where the initial majority reaches consensus. However, for this specific structure, near $m_0 = 0$ and $T = 1/2$, the ordered state is not reached for any threshold value. Furthermore, comparing with Fig. \ref{LAT_PDAGING}b with the results from the model without aging (Fig. \ref{LAT_PD}b), the discontinuous jump at $m_0 = 0$ for $T = 3/8, 4/8$ is replaced  by a continuous transition, where a range of states with $0 < |m_f| < 1$ are present around $m_0 = 0$. To determine whether these states belong to Phase ${\rm I}^{*}$, II or III, we need again a characterization of phases in terms of dynamical properties. According to the results in Figure \ref{fig:evolution_lattice_aging}, we find here the same regimes identified for random networks:
\begin{itemize}
    \item \textbf{Initial mixing regime (Phase ${\rm {\bf I}}^{*}$):}  After the initial disordering stage, the average interface density shows a very slow decay reflecting the slow growth of spatial domains in each binary state. The persistence in this phase shows a power-law decay $p(t) \sim t^{-1}$ (see $T = 1/8,2/8$ in Fig. \ref{fig:evolution_lattice_aging});
    \item \textbf{Ordered regime (Phase II):} It is characterized by coarsening dynamics that end in the absorbing states $m_f = \pm 1$. The form of the decay of the interface density depends on the value of $m_0$ (see $T = 3/8,4/8$ in Fig. \ref{fig:evolution_lattice_aging});
    \item \textbf{Frozen regime Phase III):} It characterizes by an initial tendency to order but the system very fast reaches an absorbing frozen configuration (see $T = 5/8,7/8$ in Fig. \ref{fig:evolution_lattice_aging}).
\end{itemize}

The implications of aging become explicit by comparing the dynamical properties of the cases with aging (Figure \ref{fig:evolution_lattice_aging}) and without aging (Figure \ref{fig:evolution_lattice}). When the threshold is $T<3/8$, Phase ${\rm I}$ is replaced by Phase ${\rm I}^{*}$ in which there is an initial disordering process very fast followed by a slow coarsening process that accelerates when we increase the threshold. Although the aging implications in this phase are similar to those observed in the ER graph, the coarsening process is slower  (see insets in Fig. \ref{fig:evolution_lattice_aging}a).

\begin{figure*}
     \centering
     \includegraphics[width=0.8\linewidth]{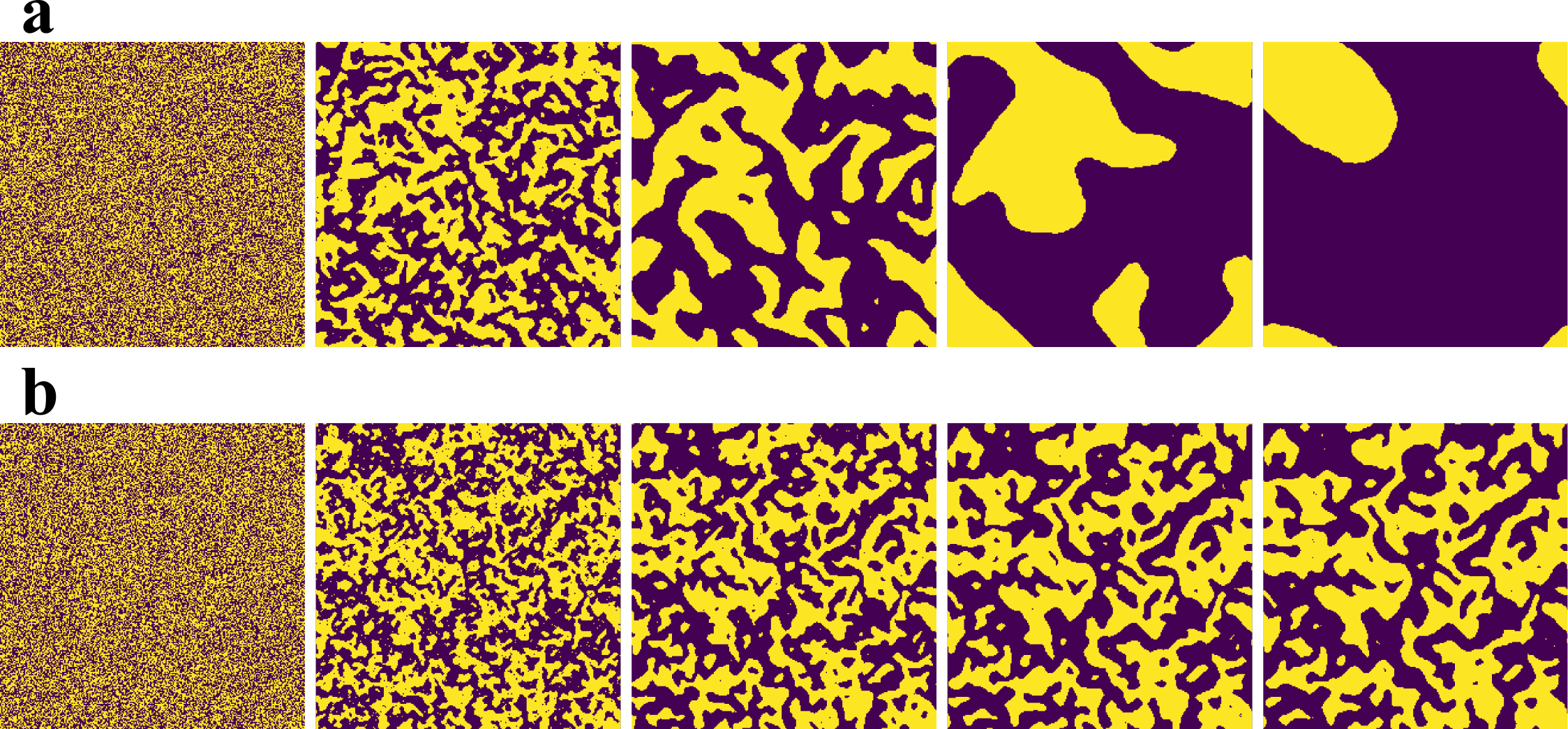}
     \caption{\label{fig:snapshots} Evolution of a single realization for $T = 0.5$ and $m_0 = 0$ using the Symmetrical threshold model (a) and the version with aging (b). Snapshots are taken after $1,10,60,440$ and $3300$ time steps in (a) and after $1,60,3300,2 \cdot 10^5$ and $5 \cdot 10^6$ time steps in (b), increasing from left to right. System size is fixed to $N = L \times L$, $L = 256$.}
\end{figure*}

In Phase II ($T=3/8, 4/8$) and when $m_0=0.5$, the system exhibits coarsening towards the ordered state $m_f=\pm 1$. In this case, the exponential decay $\rho \sim \exp(-\alpha \, t)$ observed in the absence of aging is replaced, due to aging, by a power law  $\rho \sim t^{-\alpha}$ as  noted in Ref. \cite{Abella-2022-AME}. We find $\alpha=0.5$ and $0.8$ for $T=3/8$ and $4/8$, respectively. For $m_0=0$, the power law decay of the interface density vanishes with aging, and the system exhibits a coarsening dynamics much slower than for an unbalanced initial condition. In this region of the phase diagram, spatial clusters start to grow from the initial condition, but once formed, it takes a long time for the system to reach the absorbing state $m_f = \pm 1$. 
We note that for these parameter values, the system is not able to reach $|m|$ over $0.1$ even after $10^6$ time steps, but since there is coarsening from the initial condition, the expected stationary state as $t \to \infty$ is $m_f=\pm1$. There is neither initial disordering nor freezing, these values correspond to the defined Phase II, even though the system exhibits ``long-lived segregation'' in a long transient dynamics ( see the difference with the dynamics of the model without aging in Fig. \ref{fig:snapshots}). In Fig. \ref{LAT_PDAGING}a, we differentiate Phase II from Phase III by analyzing the activity in the system: If agents are changing, even though the interface decay is slow, the system is in the Phase II. While if agents are frozen, it lies in Phase III. When comparing the ordered-frozen critical line to the one from the original model (purple line), we notice that aging causes certain values ($m_0$, $T$) that were previously in Phase II near the critical line to enter the frozen phase.

Finally, it should be noted that in Phase ${\rm I}^{*}$, the initial disordering dynamics drive the system towards $m=0$. Therefore, the subsequent coarsening dynamics follow the slow interface decay observed in Phase II for $m_0 \sim 0$. Thus, the presence of aging implies that the system asymptotically orders for any initial condition, but due to the initial disordering, the coarsening dynamics fall into the ``long-lived segregation'' regime independently of the initial condition.

\section{\label{Summary and conclusions} Summary and conclusions}

In this work, we have studied with Monte Carlo numerical simulations and analytical calculations the Symmetrical Threshold Model. In this model, the agents, nodes of a contact  network, can be in one of the two symmetric states $\pm 1$.  System dynamics follows a complex contagion process in which a node changes state when the fraction of neighboring nodes in the opposite state is above a given threshold $T$. For $T=1/2$, the model reduces to a majority rule or the zero temperature Spin Flip Kinetic Ising Model. When the change of state is only possible in one direction, say from $1$ to $-1$, it reduces to the Granovetter-Watts Threshold model \cite{granovetter-1978,watts-2002,Abella-2022-AME}.  We have considered the cases of a fully connected network, Erd\H{o}s-Rényi, and random regular networks, as well as a regular two-dimensional Moore lattice. 

We have found that, in the parameter space of threshold $T$ and initial magnetization $m_0$, the model exhibits three distinct phases, namely Phase ${\rm I}$ or mixed, Phase ${\rm II}$ or ordered, and Phase ${\rm III}$ or frozen. The existence of these three phases is robust for different network structures.
These phases are well characterized by the final state ($m_f$), and by dynamical properties such as the interface density $\rho(t)$, time-dependent average magnetization $m(t)$, persistence times $p(t)$, and mean internal time $\bar{\tau}(t)$. These phases can be obtained analytically in the mean-field case of a fully connected network. For the random networks considered, we derive an approximate master equation (AME) \cite{gleeson-2013,Abella-2022-AME} considering agents in each state according to their degree $k$,  neighbors in state $-1$, $m$, and age $j$. From this AME, we have also derived a heterogeneous mean-field (HMF) approximation. While the AME reproduces with great accuracy the results of Monte Carlo numerical simulations of the model (both static and dynamic), the HMF shows an important lack of agreement, highlighting the importance of high-accuracy methods necessary for threshold models. 


Aging is incorporated in the model as a decreasing probability to modify the state as the time already spent by the agent in that state increases. The key finding is that the mixed phase (Phase ${\rm I}$), characterized by an asymptotically disordered dynamically active state, does not always exist: the aging mechanism can drive the system to an asymptotic absorbing ordered state, regardless of how low the threshold $T$ is set. A similar effect of aging was already described for the Schelling model in Ref. \cite{Abella-2022}. When the dynamics are examined in detail, a new Phase ${\rm I}^{*}$, defined in terms of dynamical properties, emerges in the domain of parameters where the model without aging displays Phase ${\rm I}$. This phase is characterized by an initial disordering regime ($m \to 0$) followed by a slow ordering dynamics, driving the system toward the ordered absorbing states (including the one with spins opposite to the majoritarian initial option). This result is counter-intuitive since aging incorporates memory into the system, yet in this phase, the system ``forgets'' its initial state. The network structure plays an important role in the emergence of Phase ${\rm I}^{*}$ since it does not exist for complete graphs. A detailed analysis reveals that Phase ${\rm I}^{*}$ replaces Phase ${\rm I}$ only for sparse networks, including the case of the Moore lattice. For ER networks we find that, as the mean degree increases, Phase ${\rm I}$ reappears and there is a range of values of the mean degree for which phases ${\rm I}$ and ${\rm I}^{*}$ coexist. Beyond a critical value of the mean degree, Phase ${\rm I}$ extends over the entire domain of parameters where Phase ${\rm I}^{*}$ was observed.

While aging favors reaching an asymptotic absorbing ordered state for low values of $T$ (Phase ${\rm I}$), in Phase II the ordering dynamics are slowed down by aging, changing, both in random networks and in the Moore lattice, the exponential decay of the interface density by a power law decay with the same exponent. The aging mechanism is found not to be important in the frozen Phase ${\rm III}$. All these effects of aging in the three phases are well reproduced for random networks by the AME derived in this work, which is general for any chosen activation probability $p_A (j)$.

For the Moore lattice, we have also considered in detail the special case of the initial condition $m_0=0$. In this case also Phase ${\rm I}^{*}$ emerges and Phase ${\rm III}$ is robust against aging effects. However, in Phase ${\rm II}$ aging destroys the characteristic power law decay of the interface density $\rho(t) \sim at^{-1/2}$ associated with curvature reduction of domain walls. This would be a main effect of aging in the dynamics of the phase transition for the zero temperature spin flip Kinetic Ising model \cite{gunton1983}.

As a final remark on the general effects of aging in different models of collective behavior, we note that the replacement of a dynamically active disordered stationary phase by a dynamically ordering phase is generic. In this paper, we find the replacement of  Phase ${\rm I}$ by Phase ${\rm I}^{*}$. Likewise in the Voter model, aging destroys long-lived dynamically active states characterized by a constant value of the average interface density, and it give rise to coarsening dynamics with a power law decay of the average interface density \cite{fernandez-gracia-2011}. In the same way, in the Schelling segregation model, a dynamically active mixed phase is replaced, due to the aging effect, by an ordering phase with segregation in two main clusters. 
Another aging effect that seems generic, in phases in which the system orders when there is no aging, is the replacement of dynamical exponential laws by power laws. This is what happens here in  Phase ${\rm II}$ for the decay of the average interface density but, likewise, exponential cascades in the Granovetter-Watts model are replaced due to aging by a power-law growth with the same exponent \cite{Abella-2022-AME}.

Further work with the general AME used in this work would include a new approach considering the master equation, as in Ref. \cite{peralta-2020B}, in order to incorporate  finite size effects (relevant close to $m_0 = 0$) and to give a mathematical framework to describe the results in Ref. \cite{Konstantin}.

\begin{acknowledgments}

Financial support has been received from the Agencia Estatal de Investigación (AEI, MCI, Spain) MCIN/AEI/10.13039/501100011033 and Fondo Europeo de Desarrollo Regional (FEDER, UE) under Project APASOS (PID2021-122256NB-C21 and PID2021-122256NB-C22) and the María de Maeztu Program for units of Excellence, grant CEX2021-001164-M.)


\end{acknowledgments}

\appendix

\section{\label{appendix} Heterogeneous mean-field (HMF)}

When the transition and aging probabilities do not depend on $j$, $T^{\pm}_{k,m,j} = T^{\pm}_{k,m}$ and $A^{\pm}_{k,m,j} = A^{\pm}_{k,m}$, if we are not interested in the solutions $x^{\pm}_{k,m,j} (t)$ and we just want the final magnetization, Eq. \ref{eq:AME_age} is reduced to Gleeson's AME \cite{gleeson-2013} by summing variable $j$. This is a system of $(k_{\rm max}+1)(k_{\rm max}+1)$ differential equations without loss of accuracy. 

Moreover, following the steps in Ref. \cite{gleeson-2013}, we perform a heterogeneous mean-field approximation (HMF) to reduce our system to $k_{\rm max}+1$ differential equations:
\begin{align}
\frac{d}{d t} x^{-}_{k}= &- x^{-}_{k} \sum_{m=0}^{k} T^{-}_{k, m} B_{k, m}[\omega]\nonumber\\
&+\left(1-x^{-}_{k}\right) \sum_{m=0}^{k} T^{+}_{k, m} B_{k, m}[\omega],
\label{eq:HMF}
\end{align}
where $x^{-}_{k} = \sum_{j} \sum_{m}^{k} x^{-}_{k,m,j}$ and $\omega= \sum_k p_k \frac{k}{z} x^{-}_{k}$. This system of differential equations, coupled via $\omega$, cannot be solved analytically. Solving numerically with standard methods, HMF predicts a mixed-ordered transition line that qualitatively captures the critical line dependence but quantitatively differs from the numerical simulations (see the red dashed line in Figs. \ref{ER_REG_PD}a and \ref{ER_REG_PD}b and the dotted colored lines in Fig. \ref{ER_REG_PD}c). Moreover, this approximation does not predict a frozen phase in any of the networks considered. Instead, for high values of $T$, the integrated stationary solutions are always $m_f = \pm 1$, regardless of $m_0$. From this analysis, we conclude that we need sophisticated methods beyond an HMF description to describe the Symmetrical Threshold model's phase diagram, as occurs for the asymmetrical Granovetter-Watts' Threshold model (see Ref. \cite{Abella-2022-AME}).

\section{\label{appendix_HMFA} Derivation of the stationary solution via the Heterogeneous mean-field considering the age (HMFA)}

Setting the time derivatives to 0 in Eqs. \eqref{eq:HMFaging2}, we obtain the relations for the stationary state:

\begin{align}
     x^{\pm}_{k,0} = & \sum_{j=0}^{\infty} x^{\mp}_{k,j} \,  \omega_{k,j}^{\mp} \nonumber \\
     x^{\pm}_{k,j} = & x^{\pm}_{k,j-1} \, ( 1 - \omega_{k,j-1}^{\pm})  \qquad j > 0, \label{eq:SSaging4}
\end{align}
from where we extract the necessary condition $x^{-}_{k,0} = x^{+}_{k,0}$, as in Ref. \cite{chen-2020}. Notice that by setting $p_A(j) = 1$ and summing all ages $j$, we recover the HMF approximation (Eq. \ref{eq:HMF}) for the model without aging. Defining $x^{\pm}_{j}(t)$ as the fraction of agents in state $\pm 1$ with age $j$:
\begin{equation}
    x^{\pm}_{j} = \sum_k p_k \, x^{\pm}_{k,j},
\end{equation}
and placing the degree distribution of a complete graph $p_k = \delta(k-N+1)$ (where $\delta(\cdot)$ is the Dirac delta), we sum variable $k$ and rewrite Eq. \eqref{eq:SSaging4} in terms of $x^{\pm}_{j}$:
\begin{align}
     x^{\pm}_{0} = & \sum_{j=0}^{\infty} x^{\mp}_{j} \, \omega_{j}^{\mp}, \nonumber \\
     x^{\pm}_{j} = & x^{\pm}_{j-1} \, ( 1 - \omega_{j-1}^{\pm})  \qquad j > 0,   \label{eq:SSSaging2}
\end{align}
where $\omega_{j}^{\pm} \equiv \omega_{N-1,j}^{\pm}$. We compute the solution $x^{\pm}_{j}$ recursively as a function of $x^{\pm}_0$:
\begin{equation}
    x^{\pm}_{j} = x^{\pm}_0 \, F_j^{\pm} \qquad {\rm where} \qquad F_j^{\pm} = \prod_{a = 0}^{j-1} (1 - \omega_a^{\pm}),
\end{equation}
and summing all $j$,
\begin{equation}
    x^{\pm} = x^{\pm}_0 \, F^{\pm}  \qquad {\rm where} \qquad F^{\pm} = 1 + \sum_{j=1}^{\infty} F_j^{\pm}.
\end{equation}

Using the stationary condition $x^{-}_0 = x^{+}_0$, we reach:
\begin{equation}
    \frac{x^{+}}{x^{-}} = \frac{F^{+}}{F^{-}}.
\end{equation}

Notice that, for the complete graph, $\tilde{x}^{+} = x$, $\tilde{x}^{-} = 1 - x$. Therefore, $F^{\pm}$ is a function of the variable $x^{\mp}$ ($F^{+} = F(1 - x)$). Thus, we rewrite the previous expression just in terms of the variable $x$:
\begin{equation}
    \frac{x}{1- x} = \frac{F(1 - x)}{F(x)}.
\end{equation}

\section{\label{appendix_RR} Internal time recursive relation in Phase ${\rm {\bf I}}$/${\rm {\bf I}}^{*}$}

In Phase I and ${\rm I}^{*}$, the exceeding threshold condition ($m/k > T$) is full-filled for almost all agents in the system. Thus, agents will change their state and reset the internal time once activated. For the original model, all agents are activated once in a time step on average, but for the model with aging, the activation probability plays an important role. We consider here a set of $N$ agents that are activated randomly with an activation probability $p_A(j)$ and, once activated, they reset their internal time. Being $n_i(t)$ the fraction of agents with internal time $i$ at the time step $t$, we build a recursive relation for the previously described dynamics in terms of variables $i$ and $t$:

\begin{align}
     & n_1(t) = \sum_{i=1}^{t-1} p_A(i) \, n_i(t-1) \nonumber\\
     & n_i(t) = (1 - p_A(i-1) ) \, n_{i-1}(t-1)  \qquad i > 1. \label{eq:RR1}
\end{align}

This recursion relation can be solved numerically from the initial condition ($n_1(0) = 1$, $n_i(0) = 0$ for $i > 1$). To obtain the mean internal time at time $t$, we just need to compute the following:

\begin{equation}
    \label{eq:RR}
    \bar{\tau}(t) = \sum_{i=1}^{t} i \, n_i(t).
\end{equation}

The solution from this recursive relation describes the mean internal time dynamics with great agreement with the numerical simulations performed at Phase I (for the complete graph) and Phase ${\rm  I}^{*}$ (for the Erd\H{o}s-R\'enyi and Moore lattice).

\section{\label{RR_phase_diagram} Symmetrical Threshold model with aging in Random-Regular graphs}

\begin{figure}[t]
     \includegraphics[width=\columnwidth]{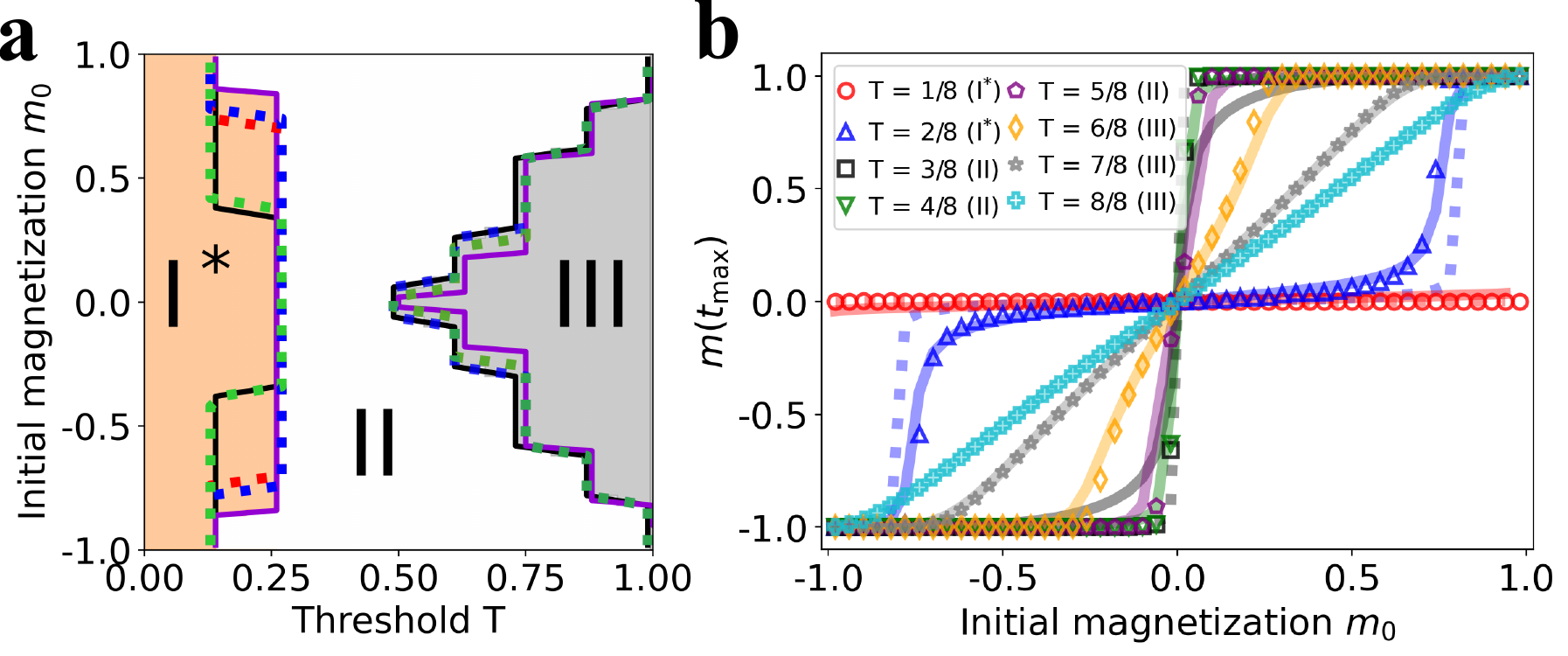}
     \caption{\label{REG_PDAGING} Phase diagram of the Symmetrical Threshold with aging model in a RR graph (a) of $N = 4 \cdot 10^4$ nodes and $\langle k \rangle = 8$. The blue, red, and green dotted lines show the borders of Phase II (first and last value of $T$ where the system reaches the absorbing ordered state for each $m_0$) computed from numerical simulations evolving until $t_{\rm max} = 10^3$, $10^4$ and $10^5$ time steps, respectively. Black solid lines show AME solution integrated $10^5$ time steps. Phase ${\rm I}^{*}$, II and III correspond with the orange, white and gray areas, respectively. The solid purple lines are the mixed-ordered and ordered-frozen critical lines for the non-aging version of the model. (b) Average magnetization at time $t_{\rm max}$ ($m(t_{\rm max})$) as a function of the initial magnetization $m_0$ for different values of the threshold $T$ (indicated with different colors and markers) in an 8-regular graph of $N = 4 \cdot 10^4$. Average performed over 5000 realizations evolved until $t_{\rm max} = 10^4$ time steps. Dotted and solid lines are the HMFA (for $T = 1/8 - 4/8$) and AME (for all $T$) solutions integrated numerically $10^4$ time steps.}
\end{figure}

Fig. \ref{REG_PDAGING} shows the borders of Phase II (first and last value of $T$ where the system reaches the absorbing ordered state for each $m_0$) obtained from Monte Carlo simulations running up to a maximum time $t_{\rm max}$ (dotted colored lines) for a RR graph. Reaching the stationary state in this model requires a large number of steps and it has a high computational cost. The two borders of Phase II exhibit different behavior as we increase the maximum number of time steps $t_{\rm max}$: while the ordered-frozen border does not change with different $t_{\rm max}$, the mixed-ordered border is shifted to lower values of $T$ as we increase the simulation time cutoff $t_{\rm max}$. As it occurs for the results in ER graphs (Fig. \ref{ER_REG_PDAGING}), our results suggest that Phase I is actually replaced in a good part of the phase diagram by an ordered phase in which the absorbing state $m_f = \pm 1$ is reached after a large number of time steps. The ordered-frozen border is now slightly shifted to lower values of the threshold $T$ due to aging. Figure \ref{REG_PDAGING}b shows the average magnetization on RR graphs with simulations running up to a time $t_{\rm max} = 10^4$. Upon comparison with Figure \ref{ER_REG_PD}c, the dependence on $m_0$ is quite similar, indicating the persistence of a transient mixed phase. This calls for a characterization of different phases in terms of dynamical properties and not only by the asymptotic value of the magnetization.

Regarding to the AME integrated solutions, Figure \ref{REG_PDAGING} shows the mixed-ordered and ordered-frozen transition lines predicted by the integration of the AME equations until a time cutoff $t_{\rm max}$, which show a good agreement with the numerical simulations. Figure \ref{REG_PDAGING}b also shows the predicted dependence of $m_f(m_0)$ for the RR graph. For comparison purposes, the numerical integration is computed until the highest $t_{\rm max}$ used in the Monte Carlo simulations. In addition, we apply the previously introduced HMFA to these random networks by numerically integrating Eqs. \eqref{eq:HMFaging2}. The results, displayed as dotted colored lines in Figure \ref{REG_PDAGING}b, show similarity to the AME solution for $T < 0.5$. Nevertheless, as it occurred for the HMF in the original model, this mathematical framework is not able to describe the frozen phase.

\section{\label{sec:temporal_dynamics} Temporal dynamics in the Symmetrical Threshold model with aging}

\begin{figure}[t]
     \centering
     \includegraphics[width=0.65\columnwidth]{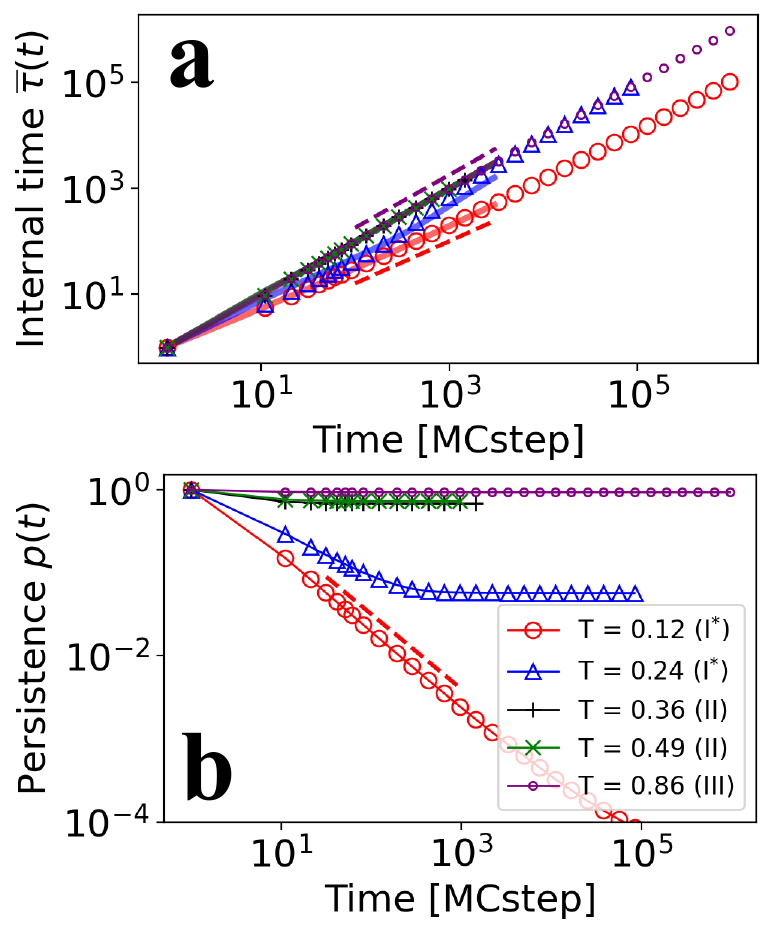}
     \caption{\label{fig:evolution_temporal_aging} Evolution of the mean internal time $\bar{\tau}(t)$ (a) and the persistence $p(t)$ (b) for the Symmetrical Threshold model with aging. The average is computed over $5000$ surviving trajectories (simulations stop when the system reaches the absorbing ordered states) for different values of $T$, shown by different markers and colors: red ($T = 0.12$) and blue ($T = 0.24$) belong to Phase ${\rm I}^{*}$, green ($T = 0.36$) and grey ($T = 0.49$) belong to Phase II and purple ($T = 0.86$) belong to Phase III. Solid colored lines are the AME integrated solutions for $10^4$ time steps, using Eq. \ref{eq:time}. The initial magnetization is $m_0 = 0.5$. The system is on an Erd\H{o}s-R\'enyi graph with $N = 4 \cdot 10^4$ and mean degree $\langle k \rangle = 8$. The dashed lines in (a) show $\bar{\tau}(t) = t$ (purple) and the solution from the recursive relation in Eq. \eqref{eq:RR} (red). The dashed red line in (b) shows $p(t) = t^{-1}$.}
\end{figure}

Fig. \ref{fig:evolution_temporal_aging} shows the evolution of the temporal dynamics via the mean internal time and the persistence. The persistence in Phase ${\rm I}^{*}$ shows a power-law decay, where $p(t)$ scales as $t^{-1}$, and the internal time shows an increase following the recursive relation given in Equation \eqref{eq:RR}, as it occurred for the mean-field scenario (Fig. \ref{fig:COM_AGING}). On the other hand, in Phase II, the persistence decays from $1$ to the fraction of nodes of the initial majority (the one that does not change state and reaches consensus) and the mean internal time scales linearly with time, $\bar{\tau}(t) \sim t$. For the internal time, the AME integrated solutions exhibit a remarkable concordance with the numerical simulations. Minor discrepancies between the numerical simulations and the integrated solutions can be attributed to the assumption of an infinitely sized system in the AME. As it occurred for the model without aging, the persistence cannot be predicted by this framework.

\bibliography{apssamp}

\end{document}